\documentclass[review]{elsarticle}
\usepackage{fullpage}
\usepackage{graphicx, float}
\usepackage{subcaption}
\usepackage{caption}
\usepackage{indentfirst}
\usepackage{siunitx}
\usepackage{amsmath}
\usepackage{gensymb}
\usepackage{booktabs}
\usepackage{multirow}
\usepackage{commath}
\usepackage{enumerate}
\usepackage{enumitem}
\usepackage{nomencl}
\usepackage{amssymb}
\usepackage{etoolbox}
\usepackage{hyperref}
\usepackage{cleveref}
\makenomenclature

\usepackage{xcolor}
\usepackage{multirow}
\usepackage[flushleft]{threeparttable}
\usepackage{textcomp}
\usepackage{gensymb}
\usepackage[export]{adjustbox}
\usepackage{tikz}
\usepackage{tablefootnote}
\usepackage{amsmath}
\usepackage[acronym,style=long3col]{glossaries}
\usepackage{chemfig}
\usepackage{changes}

\usepackage{xcolor, soul}
\sethlcolor{red}
\usepackage{lineno}

\biboptions{sort&compress}
\begin{document}

\begin{frontmatter}

\title {Numerical Investigations on Bitumen Liberation from Oil Sands Lump}


\author[mainaddress]{Mohammadhossein Golchin}

\author[mainaddress]{Siyu Chen}

\author[secondaddress]{Payman Esmaeili}

\author[mainaddress]{Petr Nikrityuk\corref{cor1}}
\ead{nikrityu@ualberta.ca}

\author[mainaddress]{Xuehua Zhang\corref{cor1}}
\ead{xuehua.zhang@ualberta.ca}

\address[mainaddress]{Department of Chemical and Materials Engineering, University of Alberta, Alberta T6G 1H9, Canada}

\address[secondaddress]{Imperial Oil Limited, Calgary, Alberta T2C 4P3, Canada}

\begin{abstract}

The dynamic liberation and digestion of viscous oil–solid aggregates under turbulent flow is a critical transport phenomenon in multiphase chemical processes, including slurry transport, soil remediation, and industrial extraction. A primary industrial application of this phenomenon is oil sands hydrotransport, where pipelines are widely utilized to simultaneously transport and extract bitumen from lump ore. However, optimizing operating conditions requires a fundamental understanding of the localized hydrodynamic mechanisms governing detachment. To address this, an three-dimensional (3D) Eulerian-Eulerian (E-E) CFD approach coupled with $k-\omega$ SST turbulence model has been utilized to investigate bitumen detachment dynamics from complex solid matrices. Six numerical cases were analyzed across lump diameters ($D$) of 4 and 5 cm and particle Reynolds numbers ($Re_p$) between 2000 and 4000. The simulation results reveal that bitumen liberation is a shear-driven process, occurring only when the shear stresses exerted by the flow on the lump surface exceed a critical threshold value. The bitumen liberation rate increases with higher $Re_p$ and smaller $D$, due to enhanced local shear forces and reduced particle shielding.  Based on these mechanistic insights, a novel semi-empirical shear-driven model was formulated. The proposed model demonstrates excellent predictive accuracy when validated against experimental liberation data, with an error less than 8.3\%. Overall, these findings clarify the fluid–particle interactions governing bitumen detachment and offer a robust, physics-based tool for predicting recovery efficiency in industrial hydrotransport operations.

\end{abstract}

\end{frontmatter}

\newpage
\section*{Highlights}
\begin{itemize}
    \item A CFD model coupled with $k-\omega$ SST turbulence is established for bitumen liberation from oil sands lump.

    \item Particle Reynolds number $Re_p$ and lump diameter affect the bitumen liberation rate.
    
    \item A shear-driven semi-empirical model is developed for predicting bitumen liberation rate.

    \item The proposed model has 8.3\% error against industrial experimental data.
\end{itemize}

\clearpage

\section{Introduction}

The detachment and digestion of highly viscous liquid films from porous solid matrices subjected to turbulent shear flows represent a complex multiphase transport phenomenon relevant to various chemical and environmental operations, including soil remediation, slurry pipeline transport, and industrial extraction processes. A primary industrial realization of this phenomenon occurs during oil sands hydrotransport, where mined ore - a heterogeneous composite of bitumen, water, and silica sands - is conditioned within pipelines \cite{masliyah2004understanding}. Under high-Reynolds-number pipe flows, intense hydrodynamic shear forces, vortex interactions, and inter-particle collisions drive the liberation of bitumen from the solid matrix \cite{wallwork2004processibility}. Understanding the fundamental fluid–particle mechanisms that govern this detachment process is critical for predicting extraction rate, reducing energy consumption, and optimizing overall recovery efficiency \cite{masliyah2011handbook,pazouki2019ablation}.

Experimental studies have been performed on reducing bitumen-particle interaction to facilitate bitumen liberation. It has been shown that thermal or chemical methods, such as utilizing hot water \cite{xia2022review}, adding surfactants \cite{wang2021understanding,zhou2018role,liu2003studies} or adjusting the pH value of the environment can help liberation by reducing bitumen viscosity or weakening the interfacial binding forces between the bitumen and the lump. For example, Dai and Chung \cite{dai1995bitumen} found that increase in both alkalinity and temperature can significantly increase the repulsive force between silica sand and bitumen, especially for coarse particles. Srinivasa et al. \cite{srinivasa2012study} designed a cell that allow real-time in-situ visualization of the bitumen liberation processes subjected to liquid flow and found that aging of the ores, excessive amount of fines and presence of salt can hinder the bitumen liberation process. Li et al. \cite{li2023effect} investigated the effect of clay on the bitumen liberation process. They found that hydrophobic and hydrophilic clay particles hinder the bitumen liberation process, and increase in alkalinity can mitigate these adverse effects. 

The separation of bitumen from solid particles subjected to external forces has been widely investigated. For example, Smith and Van de Ven \cite{smith1985separation} modeled a pendent drop on a solid particle submerged in a medium and found that the Bond number, drop volume and the wetting characteristics are important to the stability of the three phase system. Mehrabian et al. \cite{mehrabian2016oil, mehrabian2018oil} studied an oil-coated solid sphere falling into an aqueous solution and found that the separation process is shear driven, which depends on the capillary number, $Ca$. The maximum separation efficiency is reached when $Ca\sim 1$ such that viscous shear force is comparable with the interfacial tension. In addition, there exists a critical oil layer thickness beyond which the oil layer can be separated.

The fundamental model of a spherical sand particle covered by a layer of bitumen subjected to high speed water flow has been further investigated using computational fluid dynamics (CFD). For example, Shayunusov et al. \cite{shayunusov2024dynamics} established a model using a three-dimensional (3D) Coupled Level Set and Volume of Fluid (CLSVOF) method with Adaptive Mesh Refinement (AMR) for the bitumen-sand separation process at a particle Reynolds number, $Re_p$, of 500. They found that the separation process involves multi-staged shedding and recoiling of bitumen droplets due to the vortices formed at the leeward side of the particle. The thickness of the bitumen layer, $\delta$, and $Re_p$ are the two critical factor that affect the shear force applied to the bitumen layer. They proposed a regime map based on these two parameters to determine whether bitumen layer would be moved and/or separated from the sand particle \cite{shayunusov2025shear}. 

The fundamental model has been extended to model the oil sands lump digestion in hydrotransport pipelines, which includes many oil sands ores and lumps with complicated solid matrices. For example, Eskin et al. considers the convective heat transfer and shear forces for a lump moving in a turbulent flow \cite{eskin2002model}. The bitumen layer is liberated when its surface temperature exceeds a critical threshold value due to both heat transfer and shear force and the lump shrinks. They modeled lump consumption by a two-layer flow with water and fines at top and large lumps at bottom and found that the digestion rate increases as the water temperature increases, flow velocity increases and lump size decreases \cite{eskin2003simulation}. Masliyah et al. coupled convective heat transfer at the interface with a constant critical temperature associated with bitumen's adhesive strength \cite{wallwork2004processibility,masliyah2004understanding}. Building upon these frameworks, Pazouki provided vital online strain-gauge measurements of real-time mass loss and drag forces on ablating lumps to develop and validate a refined model that formulates ablation rate as a direct function of surface shear forces and temperature-dependent bitumen viscosity.\cite{pazouki2013experimental,pazouki2019ablation}

Numerical approaches are beneficial to improve the understanding of the interaction between the liquid flow and lumps. For example, Gu et al. (2019) coupled CFD with discrete element method (CFD-DEM) to simulate the temporal evolution of clay erosion in sand aggregate subjected to seepage flow \cite{gu2019based}. They identified two different erosion patterns at low and high pressure gradient. Lu et al. applied a ghost-cell based immersed boundary method with direct numerical simulation to model fluid flow and mass transfer dynamics in complex particle clusters subjected to fluid flow \cite{lu2018direct}. They showed that particle clustering restricts flow penetration, creates local concentration gradients, and reduces overall mass transfer efficiency. However, there is a lack of understanding on the separation of highly viscous fluid from complicated solid matrices subjected to turbulent water flow, which is a typical scenario for oil sands lump consumption in the industry. 

To address this knowledge gap, the present work establishes a Eulerian-Eulerian (E-E) approach coupled with the k-$\omega$ turbulent model to investigate the transient dynamics of bitumen liberation from oil sands lumps under turbulent water flow. By resolving the local hydrodynamic flow fields surrounding multi-layered particle matrices across various lump sizes and Reynolds numbers, we show the bitumen liberation is a complicated shear-driven process. We demonstrates that such a process requires a shear stress higher than the critical value to occur, and the vortices and solid particles play important roles on the process. Based on these numerical results, we propose a semi-empirical, shear-driven model to quantify bitumen liberation rates and validate its predictive accuracy against experimental benchmark data. The paper is organized as follows: Section II provides a description to the bitumen liberation problem; Section III describes the governing equations involved in the E-E model; Section IV describes the implementation of the models; Section V introduces the experimental benchmark dataset; Section VI derives the semi-empirical bitumen liberation model and compares its prediction with the experimental data; Section VII discusses the bitumen liberation mechanism and the effect of lump size and $Re_p$ based on the simulation results and a conclusion is made in Section VIII.\\

\section{Problem description}

\begin{figure}[htbp]
    \centering
    \includegraphics[width=0.9\textwidth]{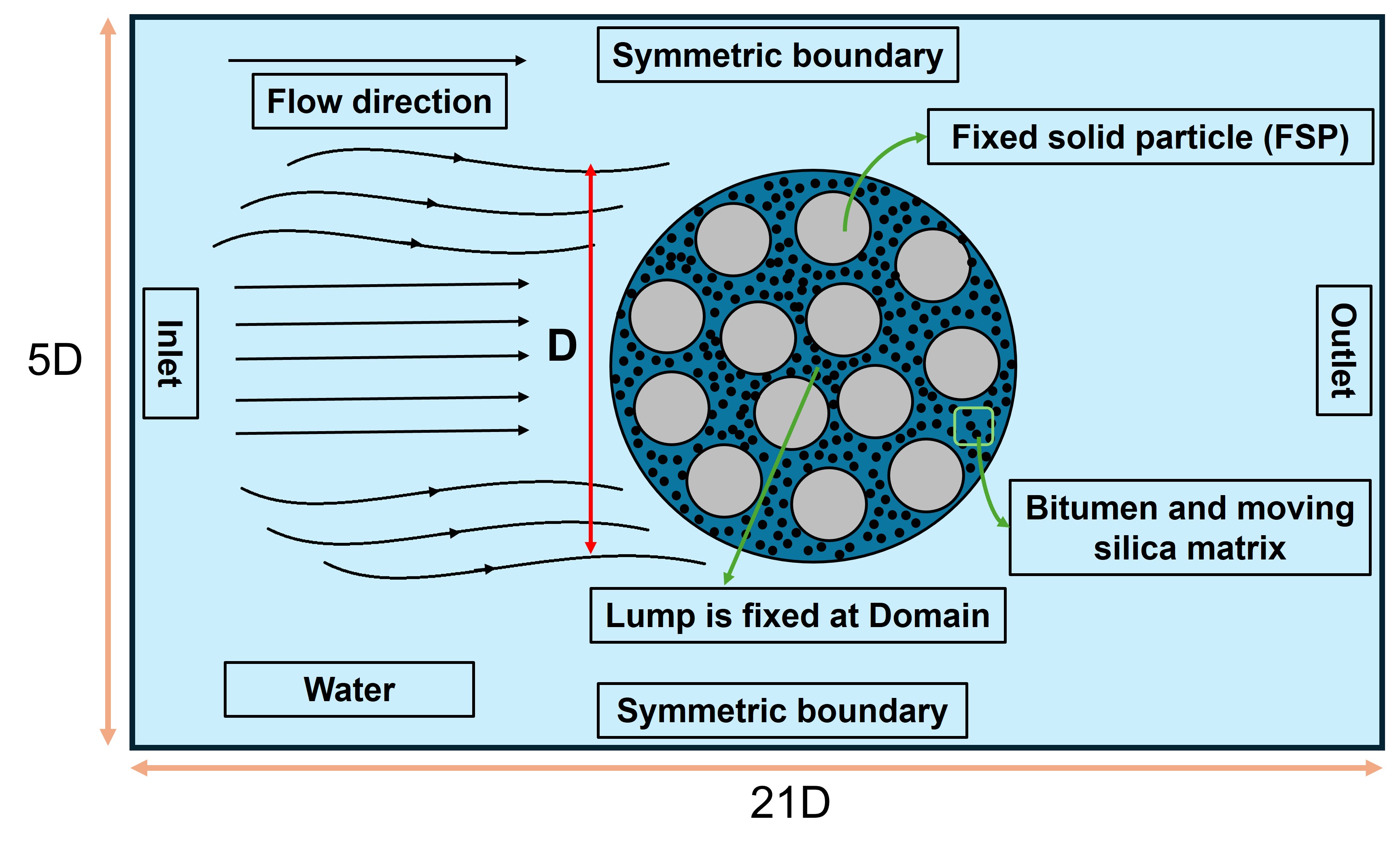}
    \caption{Schematic of the a cross-section of 3D computational domain ($5D \times 21D$) and internal structure of the fixed oil sands lump subjected to water flow. The diagram illustrates the inlet and outlet boundaries, symmetric domain boundaries, and the multi-phase composition of the lump comprising fixed solid particles (FSP) embedded within a bitumen and moving fine silica matrix.}
    \label{fig_schematic}
\end{figure}

The present work considers the bitumen liberation from an oil sands lump subjected to water flow as illustrated in Figure \ref{fig_schematic}. The oil sands lumps are considered as homogeneous spheres with monodisperse silica particles. The bitumen liberation process has been shown to be relevant to the lump Reynolds number
\begin{equation}
Re_p = \frac{\rho_w v_{rel} D}{\mu_w}
\label{eqn:relump}
\end{equation}
where $D$ is the lump diameter, $v_{rel} = |\vec{u}_{lump} - \vec{u}_w|$ is the slip velocity. As a result, $D$ and $v_{rel}$ are varied to study their effects on the bitumen liberation process. The diameter of the lump is selected to be 4 and 5 cm, which is within the range of the lump diameter in the hydrotransport pipelines (2.54 to 4 inches) \cite{masliyah2011handbook,pazouki2019ablation}. The 4 cm lump contains 16 fixed solid particles (FSP), while the 5 cm lump contains 36 FSP where each fixed solid particles have a diameter of 1 cm. The volume in the lumps within the FSP is patched with bitumen and silica sands with bitumen volume fraction of 0.33, silica volume fraction 0.6 and 0.07 volume fraction of water. The diameter of the silica sand particles is set as 123 $\mu m$, which is the characteristic diameter, $d_{50}$, of the particle size distribution in the experimental cases. Each of these two lumps are exposed to water with three slip velocities such that the $Re_p$ is 2000, 3000 and, 4000 for parametric study.

The simulation is performed in a three-dimensional computational domain with a width of 5D and a length of 21D. The dimension of the computational domain is selected to prevent outlet backflow and to ensure that the outlet boundary condition has a negligible effect on the flow field near the lump. The oil sands lump is fixed at the origin of the domain, while the surface of FSP are modeled as rigid walls. The primary phase in the simulation is the water, while the secondary phases are the bitumen and silica sands. The assumptions made for the simulation are summarized as follows:
\begin{itemize}
    \item The oil sands lump is considered as a homogeneous sphere (no shape factor considered).
    \item The silica particles are monodisperse with a size of 123 $\mu m$ (the $d_{50}$ of the PSD).
    \item The fixed solid particles are hollow regions in mesh to fix the lump in the system of coordinate.
    \item The oil sands lump is fixed at the origin of chamber.
    \item The bitumen temperature within the lump is considered as a constant homogeneous temperature of 
    $40^\circ C$. As a result, the bitumen has a constant viscosity of $\mu_B = 10 kg/m.s$ \cite{behzadfar2013viscoelastic,behzadfar2014rheology}.
    \item Bitumen and silica are distributed uniformly within the lump.
    \item The gravity is neglected due to the shear-driven property of the bitumen liberation process.

\end{itemize}

\section{Mathematical Modeling Methodology}

\subsection{Governing Equations}
The Eulerian-Eulerian approach is used to model the three phases, namely the water phase (denoted by subscript $w$), the bitumen phase (denoted by subscript $bit$), and the silica sand phase (denoted by subscript $sil$). The governing equations for the E-E model are presented in Table~\ref{table:1}. Equation (2a) is the continuity equation, which enforces the mass balance. Equations (3a), (4a) and (5a) are the momentum balance for each phase. The transient inertial forces and convective terms on the left-hand side are balanced by the pressure gradient ($\nabla P$), the viscous shear stress tensor ($\nabla \cdot \tau$), and the drag force ($\Vec{F}_{drag}$) on the right-hand side of the equations \cite{tarodiya2019numerical, brackbill1992continuum, hirt1981volume}. Silica sand is a granular material, which is modeled by the kinetic theory of granular flow (KTGF) with governing equations listed in Table \ref{table:2}. Schiller-Naumann model is used for the drag force between water and bitumen, while Gidaspow model is used for the drag forces between water and silica sand, and between silica sand and bitumen. 

The interfacial tension coefficient ($\sigma$) between water and bitumen is 0.02 N/m \cite{schramm2003influence}. The contact angle of bitumen at the rigid walls is an important parameter to determine bitumen separation from the wall. It is assumed to be $150^\circ$ in the present study \cite{shayunusov2024dynamics,shayunusov2025shear}. The bitumen behaves as a non-wetting fluid at the wall, which makes the water flow easier to separate bitumen with shear forces \cite{gao2018impact,basu1996study,masliyah2004understanding}.

\begin{table}
   \caption{Governing equations of the Eulerian-Eulerian model for the three-phase system composed of water, silica sand and bitumen~\cite{Ansys,Sekhar2023POF, tarodiya2019numerical}.}
    \label{table:1}

     \begin{tabular}{l p{10cm} l}
      \hline
       {\raggedleft Continuity}    &   &   \\
        &  $ \begin{alignedat}{8}  \frac{\partial}{\partial t}(\rho_q\alpha_q) +\nabla\cdot(\rho_q\alpha_q\Vec{v}_q) = 0 , \hspace{0.5cm}  (q=w,sil,bit)\end{alignedat} $ & (2a) \\
        
        & $\begin{alignedat}{8} \sum_q\alpha_q= 1\end{alignedat}$ & (2b) \\
        
       {\raggedleft Momentum balance}   &   &   \\
       
       {\raggedleft Liquid Phase }  &   &   \\
        & $\begin{alignedat}{8}\hspace{-3cm}\frac{\partial}{\partial t}(\alpha_{w}\rho_w\Vec{v}_w)+\nabla \cdot(\alpha_{w}\rho_w\Vec{v}_w\otimes\Vec{v}_w) =-\alpha_{w}\nabla P+\nabla \cdot\tau_w +\Vec{F}_{drag,w-sil}+\Vec{F}_{drag,w-bit}\end{alignedat}$ & (3a)  \\

        & $ \begin{alignedat}{8}\Vec{F}_{drag,w-sil} = \frac{3}{4} C_D \alpha_{sil}\rho_w \frac{\vert\vert\Vec{v}_w-\Vec{v}_{sil}\vert\vert}{d_p} (\Vec{v}_w-\Vec{v}_{sil})\end{alignedat}$ & (3b)\\ 

        & $\begin{alignedat}{8} \Vec{F}_{drag,w-bit}= \frac{3}{4} C_D \alpha_{bit}\rho_w \frac{\vert\vert\Vec{v}_w-\Vec{v}_{bit}\vert\vert}{d_p} (\Vec{v}_w-\Vec{v}_{bit})\end{alignedat} $ & (3c) \\

        {\raggedleft Silica Sand Phase}  &   &   \\
        & $\begin{alignedat}{8}\hspace{-3cm}\frac{\partial}{\partial t}(\alpha_{sil}\rho_{sil}\Vec{v}_{sil})+\nabla\cdot(\alpha_{sil}\rho_{sil}\Vec{v}_{sil}\otimes\Vec{v}_{sil}) =-\alpha_{sil}\nabla P-\nabla P_{sil}+\nabla \cdot\tau_{sil} \end{alignedat}$& \\
        &$\begin{alignedat}{18}+\Vec{F}_{drag,sil-w}+\Vec{F}_{drag,sil-bit}+\beta_{ij}(\Vec{v}_{sil_j}-\Vec{v}_{sil_i})\end{alignedat}$ & (4a) \\

          & $\begin{alignedat}{8}\hspace{-3cm} \beta_{ij} = \frac{3(1+e_{ij})(\frac{\pi}{2}+C_{fr,ij}\frac{\pi^2}{8})\alpha_{sil_i}\rho_{sil_i}\alpha_{sil_j}\rho_{sil_j}(d_{sil_i}+d_{sil_j})^2 g_{0,ij}}{2\pi(\rho_{sil_i}d^3_{sil_i}+\rho_{sil_j}d^3_{sil_j})}  \vert\vert\Vec{v}_{sil_i}-\Vec{v}_{sil_j} \vert\vert \end{alignedat}$ & (4b) \\
        
            & $\begin{alignedat}{8}\hspace{-2cm} \tau_{sil}=\alpha_{sil}\mu_{sil}(\nabla\Vec{v}_{sil} + (\nabla\Vec{v}_{sil})^T)+\alpha_{sil}(\lambda_{sil}-\frac{2}{3}\mu_{sil})(\nabla \cdot \Vec{v}_{sil})\Bar{\Bar{I}}\end{alignedat}$ & (4c)\\

        & $ \begin{alignedat}{8} \Vec{F}_{drag,sil-w} =-\Vec{F}_{drag,w-sil} \end{alignedat}$ &(4d) \\

        & $\begin{alignedat}{8}         \Vec{F}_{drag,sil-bit}= \frac{3}{4} C_D \alpha_{bit}\rho_{sil} \frac{\vert\vert\Vec{v}_{sil}-\Vec{v}_{bit}\vert\vert}{d_p} (\Vec{v}_{sil}-\Vec{v}_{bit})\end{alignedat} $ & (4e) \\

        {\raggedleft Bitumen Phase}  &   &   \\

        & $\begin{alignedat}{8}\hspace{-3cm}\frac{\partial}{\partial t}(\alpha_{bit}\rho_{bit}\Vec{v}_{bit})+\nabla\cdot(\alpha_{bit}\rho_{bit}\Vec{v}_{bit}\otimes\Vec{v}_{bit}) =-\alpha_{bit}\nabla P +\nabla \cdot\mu_{bit}(\nabla\Vec{v}_{bit} + \nabla\Vec{v}_{bit}^T) \end{alignedat}$& \\
        &$\begin{alignedat}{18}+\Vec{F}_{drag,bit-w}+\Vec{F}_{drag,bit-sil}\end{alignedat}$ & (5a) \\

        & $ \begin{alignedat}{8} \Vec{F}_{drag,bit-w} =-\Vec{F}_{drag,w-bit} ; \ \begin{alignedat}{8} \Vec{F}_{drag,bit-sil} =-\Vec{F}_{drag,sil-bit} \end{alignedat} \end{alignedat}$ &(5b) \\

  \\

     \hline

    \end{tabular}
\end{table}

 \begin{table}
 \begin{center}
     \centering

     \caption{Kinetic theory of granular flow equations~\cite{Lun_Savage_Jeffrey_Chepurniy_1984,gidaspowbook,ParvaEulerianmultifluid,golchin2026impacts} (\textit solid phases (silica particles)).}
\label{table:2}

 \begin{tabular}{cc}

 \hline

  $\begin{alignedat}{8}\hspace{-2cm}\lambda_{s_i}=\frac{4}{3}\alpha_{s_i}^2\rho_{s_i}d_{s_i}g_{0,ii}(1+e_{ij})(\frac{\Theta_{s_i}}{\pi})^{1/2}\end{alignedat}$ &(6a)\\
  $\begin{alignedat}{8}\hspace{-2cm}g_{0,ii}=[1-(\Sigma_{i=1}^2\frac{\alpha_{s_i}}{\alpha_{s,max}})^{1/3}]^{-1} + \frac{d_{s_i}}{2}\Sigma_{i=1}^{2}\frac{\alpha_{s_i}}{d_{s_i}}\end{alignedat}$ & (6b)\\
 $\begin{alignedat}{8}\hspace{-2cm}g_{0,ij}=\frac{d_{s_i}g_{0,ii}+d_{s_j}g_{0,jj}}{d_{s_i}+d_{s_j}}\end{alignedat}$ & (6c)\\
$\begin{alignedat}{8}\hspace{-2cm}\Theta_{s_i}=\frac{1}{3}\vert\vert\Vec{v'}_{s_i}\vert\vert^2\end{alignedat}$ & (6d)\\
 $\begin{alignedat}{8}\hspace{-2cm}0=(-P_{s_i}\Bar{\Bar{I}}+\Bar{\Bar{\tau_{s_i}}}):\nabla\Vec{v}_{s_i}-\gamma_{\Theta_{s_i}}+\phi_{l_i}\end{alignedat}$ & (6e)\\
 $\begin{alignedat}{8}\hspace{-2cm}\gamma_{\Theta_{s_i}}=\frac{12(1-e_{ii}^2g_{0,ii})}{d_{s_i}\pi^{1/2}}\rho_{s_i}\alpha_{s_i}^2\Theta_{s_i}^{3/2}\end{alignedat}$ & (6f)\\
 $\begin{alignedat}{8}\hspace{-2cm}\phi_{l_i}=-3K_{l_i}\Theta_i\end{alignedat}$ & (6g)\\
 $\begin{alignedat}{8}\hspace{-2cm}P_{s_i}=\alpha_{s_i}\rho_{s_i}\Theta_{s_i}[1+2\Sigma_{j=1}^2 (\frac{d_{s_i}+d_{s_j}}{2d_{s_i}})^3 (1+e_{ij}\alpha_{s_j}g_{0,ij}))]\end{alignedat}$ & (6h) \\
 $\begin{alignedat}{8}\hspace{-2cm}\mu_{s_i}=\mu_{s_i,col}+\mu_{s_i,kin}+\mu_{s_i,fr}\end{alignedat}$ & (6i)\\
 $\begin{alignedat}{8}\hspace{-2cm}\mu_{s_i,col}=\frac{4}{5}\alpha_{s_i}\rho_{s_i}g_{0,ii}(1+e_{ij})(\frac{\Theta_{s_i}}{\pi})^{1/2}\alpha_{s_i}\end{alignedat}$ & (6j) \\
 $\begin{alignedat}{8}\hspace{-2cm}\mu_{s_i,kin}=\frac{10\rho_{s_i}d_{s_i}(\Theta_{s_i}\pi)^{1/2}}{96\alpha_{s_i}(1+e_{ij})g_{0,ii}}[1+\frac{4}{5}g_{0,ii}\alpha_{s_i}(1+e{ii})]^2\alpha_{s_i}\end{alignedat}$ & (6k) \\
 $\begin{alignedat}{8}\hspace{-2cm}\mu_{s_i,fr}=\frac{P_{s_i}\sin\varphi_{s_i}}{2I_{2D}^{1/2}}\end{alignedat}$ & (6l) \\
 \hline

 \end{tabular}
 \end{center}
 \end{table}

\subsection{Turbulence Model and Wall Functions}

The hydrotransport in the industrial pipelines for oil sands is highly turbulent, with a Reynolds number ($Re=\rho_w U D_{pipe}/\mu_w$) on the order of $10^6$ \cite{golchin2026impacts}. As a result, turbulence models should be used to resolve the flow at the boundary layer near the wall, as well as the vortices, to accurately predict bitumen detachment. While the standard $k-\epsilon$ model \cite{JONES1972301,LAUNDER1974131} is suitable for bulk flows, the $k-\omega$ model \cite{Wilcox1988} is more accurate in predicting flow separation and adverse pressure gradients in the near-wall region. As a result, the $k-\omega$ shear stress transport (SST) model \cite{menter1993zonal,menter1994two} is employed in the present study, which blends the $k-\omega$ model in near-wall region with $k-\epsilon$ model at region far from the wall. The governing equations for the $k-\omega$ SST model are listed in Table~\ref{tab: Turbulence} \cite{Menter2003}.

Accurate resolution of near-wall flow field is important in determining the bitumen detachment from the wall. As a result, the viscous sublayer should be resolved directly by refining the mesh within the lump and near the fixed solid boundaries \cite{cant2001sb, esch2003heat,shayunusov2025shear}. The mesh used has a near-wall $y^{+}$ value between 2.7 and 4.6, which is below 5 to resolve the viscous sublayer. An automatic wall treatment, which blends a low-Reynolds-number formulation with a high-Reynolds-number logarithmic wall function, is used to model the turbulent properties at the wall boundary.

\begin{table}[!ht]
    \centering
    \caption{$k-\omega$ SST model equations \cite{Menter2003}.}
\label{tab: Turbulence}

\begin{tabular}{l l l}
\hline
\\
Turbulence kinetic energy & &
\\
$ \begin{alignedat}{8}
    \frac{\partial k}{\partial t}+ U_{j}\frac{\partial k}{\partial x_{j}}=P_{k}-\beta^*k\omega+\frac{\partial}{\partial x_{j}}\left[(\nu +\sigma_{k}\nu_{T})\frac{\partial k}{\partial x_{j}}\right]
\end{alignedat} $ & (7a) & \\

Specific dissipation rate & &\\
$ \begin{alignedat}{8}
    \frac{\partial \omega}{\partial t}+ U_{j}\frac{\partial \omega}{\partial x_{j}}=\alpha S^2-\beta \omega^2+\frac{\partial}{\partial x_{j}}\left[(\nu +\sigma_{\omega}\nu_{T})\frac{\partial \omega}{\partial x_{j}}\right]+2(1-F_{1})\sigma_{\omega2}\frac{1}{\omega}\frac{\partial k}{\partial x_{i}}\frac{\partial \omega}{\partial x_{i}}
\end{alignedat}$ & (7b) & \\
Kinematic eddy viscosity & &\\
$\begin{alignedat}{8}
    \nu_{T}=\frac{a_{1}k}{\max(a_{1}\omega,SF_{2})}
\end{alignedat}$ & (7c) & \\
Closure coefficients and auxiliary relations & &\\
$\begin{alignedat}{8}
    F_{1}=\tanh\left[\left[\min\left[\max\left(\frac{\sqrt{k}}{\beta^*\omega y},\frac{500\nu}{y^2\omega}\right),\frac{4\sigma_{\omega2}k}{CD_{k\omega}y^2}\right]\right]^4\right]\end{alignedat}$ & (7d) & \\
$\begin{alignedat}{8}
    F_{2}=\tanh\left[\left[\max\left(\frac{2\sqrt{k}}{\beta^*\omega y},\frac{500\nu}{y^2\omega}\right)\right]^2\right]\end{alignedat}$ & (7e) & \\
$\begin{alignedat}{8}
     CD_{k\omega}=\min\left(2\rho\sigma_{\omega2}\frac{1}{\omega}\frac{\partial k}{\partial x_{i}}\frac{\partial \omega}{\partial x_{i}}, 10^{-10}\right) \end{alignedat}$ & (7f) & \\
    $\begin{alignedat}{8}
    P_{k}=\min\left(\tau_{ij}\frac{\partial U_{i}}{\partial x_{j}} ,10\beta^* k\omega\right) \end{alignedat}$ & (7g) & \\
 $\begin{alignedat}{8}
    a_{1}= \frac{5}{9} , \alpha=0.44, \beta^*=0.09, \beta=0.0828, \sigma_{k}=0.85, \sigma_{\omega}=0.5, \sigma_{\omega_{2}}=0.856 \end{alignedat}$ & (7h) & \\
\hline     
\end{tabular}
\end{table}

\section{Numerical Implementation and Solution Strategy}

The numerical simulations are transient models solved using the finite volume method with wall-bounded axis-symmetric 3D computational domains. For the $5 cm$ oil sands lump with 36 FSPs, the domain has a height of $0.25 m$ and a length of $1.05 m$. For the $4 cm$ oil sands lump with 16 FSPs, the domain has a height of $0.2 m$ and length of $0.84 m$. Both domains maintain identical relative dimensions scaled directly to the lump diameter ($R = 5D$ and $L = 21D$) to ensure that the flow field are comparable across the cases. Furthermore, extending the symmetry boundaries to $5D$ radially and $21D$ axially ensures that the zero-normal-velocity constraint on the symmetry walls does not artificially confine or accelerate the approaching free-stream flow, and the outlet boundary condition has an negligible effect on the flow near the lump. 

\subsection{Boundary Conditions and Initialization}
The boundary condition at the inlet is a uniform velocity profile, where the fluid velocity ($U$) is selected based on the target $Re_p$. The outlet boundary condition is selected as outflow.  Symmetry boundary conditions were applied to all lateral perimeters of the chamber (top, bottom, left, and right sides). The contours of the FSPs are modeled as no-slip walls. Under high-pH conditions, the silica sands exhibit strong water-wettability. This behavior is captured by prescribing a static three-phase contact angle of $150^\circ$ (measured through the bitumen phase) at the fixed solid particle walls. At $t=0s$, the domain is initialized with a uniform water flow field, and the bitumen and silica sands volume fractions are patched on to the lump region. The velocity is patched to 0 m/s for material inside the lump region.

\subsection{Solver Parameters and Discretization}
The pressure-velocity coupling is resolved utilizing the coupled algorithm, which provides robust convergence for multiphase systems with severe viscosity gradients. For spatial discretization, the least squares cell-based method is utilized for computing gradients. The PRESTO! scheme is applied for pressure interpolation to manage the complex pressure fields around the lump. The convective terms in the momentum, turbulent kinetic energy, and specific dissipation rate equations are discretized using a second-order upwind scheme to minimize numerical diffusion. To maintain a sharply defined water-bitumen interface and to reduce smearing, the modified HRIC scheme is used as the volume fraction solver. The convergence criterion is $10^{-6}$ for all the residual components at each time step.

\subsection{Mesh Study and Time Step}
\begin{figure}[htbp]
    \centering
    \includegraphics[width=1\textwidth]{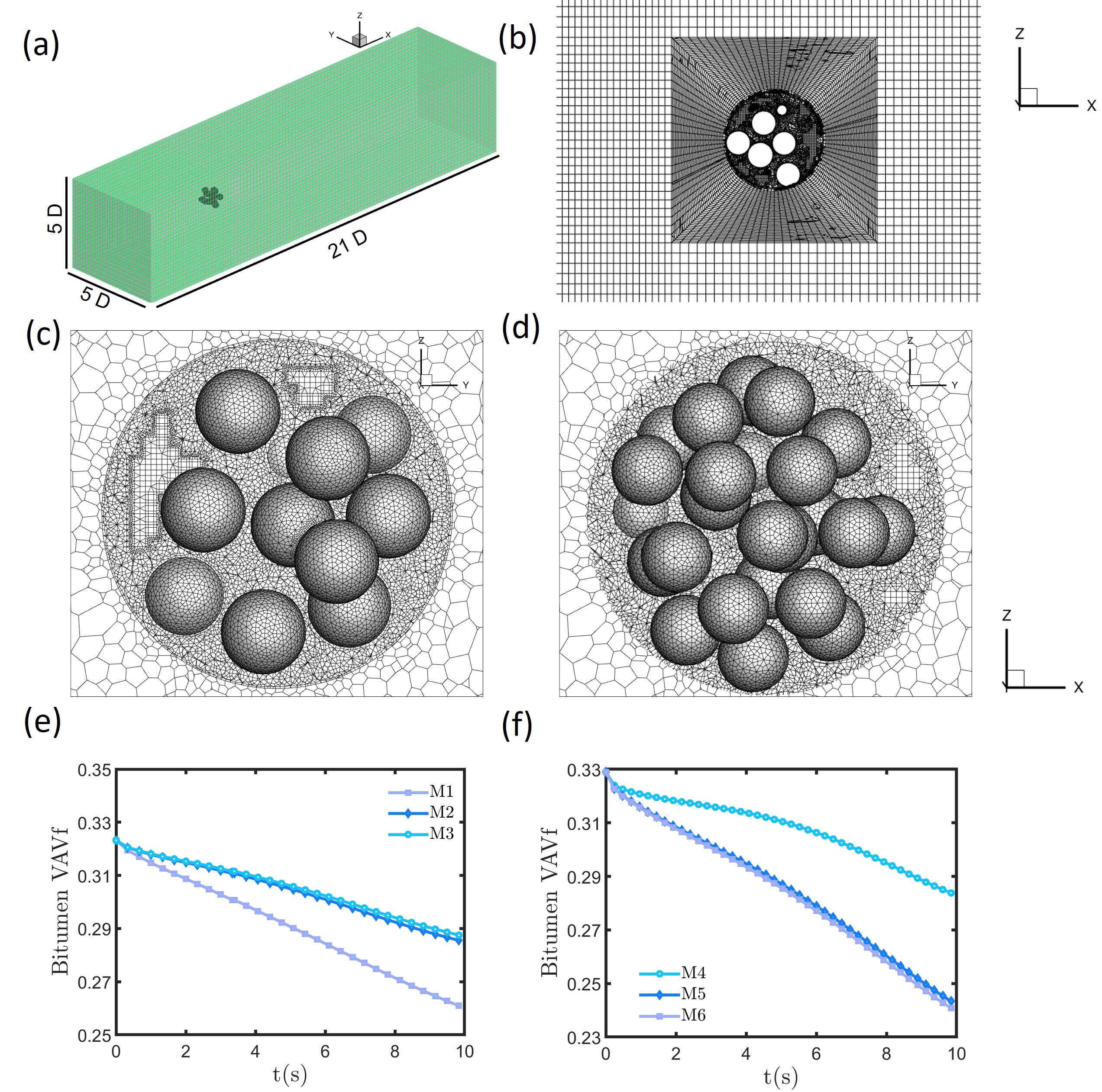}
    \caption{(a) 3D domain with dimensions respective to the lump diameter; (b) Mesh structure at the zone where lump and water interact; (c) Mesh structure inside the lump for 4 cm lump; (d) Mesh structure inside the lump for 5 cm lump; (e) Volume averaged volume fraction of bitumen versus time for 5 cm lump (M1-M3) (d)Volume averaged volume fraction of bitumen versus time for 4 cm lump (M4-M6).}
    \label{fig_mesh}
\end{figure}

A mesh dependency analysis is performed to select a proper mesh structure that balances the accuracy of numerical model with the computational cost. For the 5 cm lump case, three mesh structures are generated. Mesh M1 has 230,000 elements and mesh M2 has 920,000 elements by refining M1 within the lump. Mesh M3 has 1,079,000 elements, which is generated by further refining M2 at the solid walls within the lump. The same refinement process has been applied to the 4 cm lump case with 210,000 elements (M4), 848,000 elements (M5) and 986,000 elements (M6), respectively. Such an approach generates fine mesh structures inside the lump and at the lump interface, where the physics is most relevant to bitumen detachment, while maintaining a coarser grid in the bulk flow, as shown in Figure \ref{fig_mesh} (a-d). The performance of the mesh structures are evaluated using the volume averaged volume fraction (vavf) of bitumen. Figure \ref{fig_mesh} (e-f) show that by increasing mesh element from M2 to M3 and from M5 to M6 results in a change in vavf of bitumen less than 2\%. As a result, mesh M3, and M6 has been selected for the simulations. 

In order to ensure convergence, a sensitivity analysis is conducted on the time steps using time steps of $0.01\ s$, $0.001\ s$, and $0.0001\ s$. It is found that using the time steps of $0.001\ s$ and $0.0001\ s$ generates similar converged results. Therefore, the time step of $0.0001\ s$ is chosen. The maximum flow time is set to $10\ s$ for the bitumen liberation. A summary of the details for the solver settings is listed in Table~\ref{tab:model_setting}. 

\begin{table}[]

    \centering
    \caption{Solver settings used in CFD simulations.}
    \label{tab:model_setting}

\begin{centering}
\begin{tabular}{l l }

\hline
Model & Scheme \\
\hline
Multiphase model  & Eulerian-Eulerian (E-E)  \\
Turbulence model & $k-\omega$ SST\\
Turbulence multiphase  &  Mixture\\
Phase-Velocity Coupling & Phase Coupled\\
Gradient & Least squares cell based\\
Pressure  & PRESTO! \\
Momentum  & Second order upwind\\
Volume fraction & Modified HRIC \\
Turbulent kinetic energy & Second order upwind \\
Specific dissipation rate & Second order upwind \\
Transient formation & Second order implicit \\
Water shear condition on wall (particles)  & No-slip\\
Specularity coefficient & 0.2 \\
Restitution coefficient & 0.9 \\
Time step & 0.0001s \\
Maximum total flow time & 10 s \\
\hline

\end{tabular}
\end{centering}
\end{table}

\section{Experimental Data}
The CFD results of the bitumen liberation process are compared with the experimental data from the Saskatchewan Research Council (SRC) for validation \cite{impinternal}. The SRC experimental setup consists of a 50 m closed pipeline loop with an inner diameter of 10.16 cm (4 inches). The experiments used oil sands ores with different properties collected by Imperial Oil, Syncrude, NAIT Institution and University of Alberta. The experiments used these ores to study effect of temperature, and oil sands grade on the bitumen liberation process. The oil sands ores were crushed to small lumps with a maximum size of 1 inch and mixed in a slurry preparation tank with 60 $wt\%$ of oil sands lumps and  40 $wt\%$ of water at temperatures ranging from 37 to 50 $^\circ C$. The prepared slurry was then injected into the pipeline and circulated at a flow velocity of $2\text{ m/s}$ for an hour. Then the slurry was collected into a 100-liter barrel and the bitumen froth accumulated at the top was collected. The amounts of bitumen liberated in the froth and in the slurry was quantified by the standard Dean-Stark extraction method \cite{ma2012pyrolysis}. The bitumen recovery rate is calculated by 
\setcounter{equation}{7}
\begin{equation}
    Bitumen\ Recovery (BR) \% =\frac{Extracted\ bitumen\ from\ froth}{Extracted\ bitumen\ from\ froth\ and\ slurry} 
    \label{EQN_BR}
\end{equation}
The experiments and their conditions used for validation are listed in Table \ref{tab:experiment_table_SRC}.

\begin{table}[]

    \centering
    \caption{Operating conditions of SRC experiments used for validation}
    \label{tab:experiment_table_SRC}

\begin{centering}
\begin{tabular}{l l l l l l l }

\hline
Exp no. & T ($\degree$ C) & bubble (vol\% slurry) & $d_{50} (\mu m)$ & Silica ($wt\%$) & Water ($wt\%$) & Bitumen ($wt\%$) \\
\hline
Exp 1 & 45 & 5 & 150.6 & 86.7 & 5.2 & 8.1 \\
Exp 2  & 47 & 5 & 117.2 & 87.3 & 3.2 & 9.5\\
Exp 3  & 50 & 5 & 150.2 & 86.3 & 5.3 & 8.3 \\
Exp 4  & 37 & 5 & 148.9 & 83.8 & 5.8 & 11.5 \\

\hline

\end{tabular}
\end{centering}
\end{table}

\section{Bitumen Liberation Model in Hydrotransport Pipelines}
The bitumen liberation process in hydrotransport pipelines is driven by the shear forces. Since there is a minimum value of shear stress $\tau_{cr}$ required to separate bitumen from the solid matrix \cite{shayunusov2025shear}, an empirical correlation between the bitumen liberation rate and shear stress $\tau$ is proposed with the form
\begin{equation}
    \frac{dD}{dt} =-\alpha \left ( max \left\{0,  (\tau - \tau_{cr})^\beta \right\} \right )
\label{eq:shear}
\end{equation} 
where $\alpha$ and $\beta$ are the two fitting parameters.

\begin{figure}[htbp]
    \centering
    \includegraphics[width=1\textwidth]{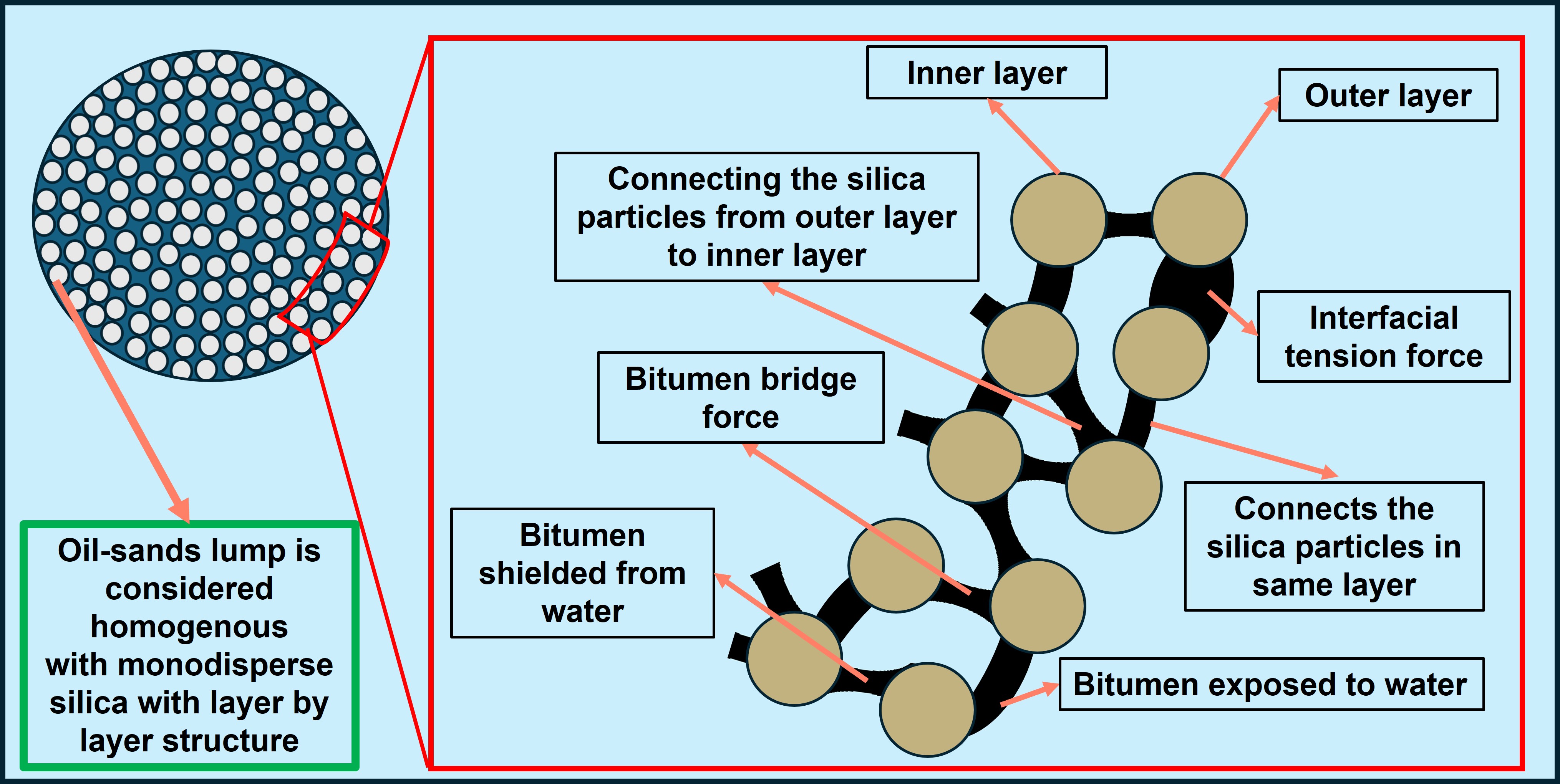}
    \caption{A schematic of an ideal spherical oil sands lump composed of bitumen and multiple layers of silica sand particles. The monodisperse silica sand particles distribute uniformly in the oil sands lump. The enlarged view shows the bitumen bridges connecting particles in the neighboring layers, as well as the bitumen covered on the particles.}
    \label{fig_bitumeninetranl}
\end{figure}

The wall shear stresses experienced at the surface of the spherical lump in the water flow can be expressed as 
\begin{equation}
\tau=\frac{1}{2}\rho_{w}C_fv_{p}^2
\end{equation}
where $C_f$ is the skin friction coefficient. For the cases in the present study, $Re_x=Re_p$ is in the range of 2000 to 4000. As a result, $C_f$ derived from the Blasius boundary layer is used\cite{schlichting2016boundary}, which is
\begin{equation}
C_f=0.664Re_p^{-0.5}
\end{equation}
The shear stress can be expressed in terms of $Re_p$ as
\begin{equation}
\tau=0.332\mu_w^2\frac{Re_p^{1.5}}{\rho_wD^2}
\end{equation}
In order to establish an expression for the critical particle Reynolds number, $Re_{cr}$, to calculate $\tau_{cr}$ in equation \ref{eq:shear}, a force balance on the outermost layer of the lump is considered. Figure \ref{fig_bitumeninetranl} shows a schematic of a spherical oil sands lump composed of bitumen and multiple layers of monodisperse silica sand particles. The silica sand particles are distributed uniformly in the lump. The bitumen not only covers the surface of silica sand particles, but also forms bridges that bind particles in neighboring layers together. In order for the water flow to peel off bitumen or sand particles from the outermost layer, the drag force inserted on the layer, $F_{drag}$, should overcome both the interfacial tension between bitumen and sand particles, $F_{\sigma}$, as well as the bitumen bridge between the outermost layer and the layer beneath it, $F_{bridge}$, i.e.,
\begin{equation} 
F_{drag} = F_{\sigma} + F_{bridge}
\label{eqn:abltion_force}
\end{equation}

The drag force of the flow on the sphere is 
\begin{equation}
    F_{Drag} = \frac{1}{8} C_D \rho_wv_{rel}^2 \pi D^2
    \label{eqn:drag_force}
\end{equation}
where $C_D$ is the drag coefficient.

The interfacial tension force between bitumen and particles on the outermost layer is
\begin{equation}
    F_{\sigma} =2 \gamma L \cos^2 \frac{\theta}{2}
    \label{eqn_single_interface}
\end{equation}
where $\gamma$ is the interfacial tension coefficient, $\theta$ is the contact angle and $L$ is the total contact length between bitumen and particles in the outermost layer. Since the diameter of sand particles is much less than the diameter of the lump, i.e. $d\ll D$, assuming that the sand particles are uniformly distributed in the outermost layer, the spatial distribution of the particles in the outermost layer can be approximated by the volume distribution of the particle in the lump, $V_{fS}$ \cite{west2012introduction}. Thus, 
\begin{equation}
    N_{ss} = V_{fS}\cdot\frac{D^2}{d^2}
    \label{eqn:number-partilces-surface}
\end{equation}
where $N_{ss}$ is the number of sand particles in the outermost layer. Since the contours of the particles may not be all in touch with bitumen, $L$ is obtained by scaling the total contact length with the volume fraction of bitumen, $V_{fB}$,
\begin{equation}
    L=\pi d V_{fB}N_{ss}
    \label{eqn_lenghth_active}
\end{equation}
By substituting equation \ref{eqn:number-partilces-surface} and \ref{eqn_lenghth_active} into equation \ref{eqn_single_interface}, the total interfacial tension force is
\begin{equation}
    F_{\sigma} =2\pi \gamma \cos^2 \frac{\theta}{2}  V_{fB}V_{fS}\frac{D^2}{d}
    \label{eqn_interface_total}
\end{equation}

Equation \ref{eqn_interface_total} indicates that $F_{\sigma}$ increases for smaller particle size or larger lump since the number of the particles in the outermost layer increases. Adjusting the surface properties by adding surfactants or pH value of the solution can increase $\theta$, which results in lower $F_{\sigma}$ and making bitumen easier to detach from the solid particles. In addition, the magnitude of $F_{\sigma}$ depends on both volume fraction of bitumen and sand particles, which are usually two competing factors. For example, in low-grade ores, while a lower bitumen volume fraction can have lower bitumen binder availability, it contains more solid particles which can potentially increase $F_{\sigma}$. As a result, the effect of bitumen and sand particle volume fractions on $F_{\sigma}$ can be complicated in real applications. 

The forces required to break a single viscous bitumen bridge between two sand particles can be obtained by the Stefan adhesion model \cite{chan2021torsional,qian2011motion}
\begin{equation}
F_{Stefan} = \frac{3\pi\mu_{B}d^{4}v_{sep}}{2h^3}
\label{eqn:stefan}
\end{equation}
where $v_{sep}$ is the velocity to separate two particles, $h$ is the distance between the two particles. For a viscoelastic bitumen bridge, it breaks when the length is greater than a critical length $h_{cr}$ \cite{read2003shell,pitois2000liquid,butler2022liquid}
\begin{equation}
h_{cr}=\left(1+\frac{\theta}{2}\right)V_{bridge}^{\frac{1}{3}}
\end{equation}
where $V_{bridge}$ is the volume of bitumen in the bridge, which can be approximated by the Fisher/Pietsch geometric approximation for a medium-sized bridge with filling angle $\varphi=24^\circ$ as \cite{fisher1926capillary,pietsch1997size,song2021lateral}
\begin{equation}
V_{bridge}=\frac{\pi}{8}d^3\tan^4{(\varphi)} \approx 0.015 d^3
\end{equation}

At the moment of detachment, the separation velocity $v_{sep}$ is approximately the water skin velocity at oil sands lump surface\cite{zhu2012particle}. In addition, for the dynamic interface, the velocity is lowered due to the viscosity difference between water and bitumen \cite{chapman1965heat,sieder1936heat}. As a result, the separation velocity can be expressed as
\begin{equation}
v_{sep} = \sqrt{\frac{C_f}{2}} v_{rel} (\frac{\mu_w}{\mu_B})^{0.14} 
\end{equation} 
 
The total number of bitumen bridge between the outermost layer and the layer beneath it can be used by the similar method as in equation \ref{eqn:number-partilces-surface} 
\begin{equation}
    N_{BS} = \frac{\pi\left(D-2l_{d}\right)^2}{\pi d^2} V_{fS}\approx V_{fS}\cdot\frac{D^2}{d^2}
    \label{eqn_bitumen_bridge number}
\end{equation}
where $N_{BS}$ is the number of available sites for bitumen bridges. $l_{d}$ is the distance between two neighboring sand particle layers and $l_d\sim d\ll D$ for a lump with closely packed sand particles \cite{gidaspow1994multiphase,syamlal1988simulation}. Similarly, the number of the total bitumen bridge is obtained by scaling $N_{BS}$ with the bitumen volume fraction, i.e.,
\begin{equation}
 N_{AS}= N_{BS}V_{fB}
 \label{eqn:nas}
\end{equation}

As a result, based on equation \ref{eqn:stefan} - \ref{eqn:nas}, the total resistance force exerted by the bitumen bridge is
\begin{equation}
F_{bridge} = F_{Stefan} N_{AS}= 6\pi \frac{\mu_{B} v_{rel} \sqrt{\frac{C_f}{2}}}{\left(1+\frac{\theta}{2}\right)^3} \left(\frac{D^2}{d}\right) V_{fB} V_{fS} (\frac{\mu_w}{\mu_B})^{0.14}
\label{eqn:bridge}
\end{equation}

Equation \ref{eqn:bridge} indicates a dependency of $F_{bridge}$ on bitumen viscosity, which changes with temperature. Furthermore, $F_{bridge}$ increases with a larger lump-to-particle size ratio ($D/d$) due to an increased number of bitumen bridges.

By substituting equation \ref{eqn:drag_force}, \ref{eqn_interface_total} and \ref{eqn:bridge} into equation \ref{eqn:abltion_force}, the critical Reynolds number $Re_{cr}=\rho_{w} v_{rel,cr}D/\mu_w$ can be obtained by solving
\begin{equation}
Re_{cr}^2 = \left( \frac{16}{C_D} \frac{\rho_w \cdot \gamma \cdot D}{ \mu_w^2}  \cdot \cos^2\frac{\theta}{2} \cdot \left( V_{fB} \cdot V_{fS} \cdot \frac{D}{d}\right)\right) + 
    \left ( \frac{48}{ C_D } \cdot \frac{\lambda_{B}^{0.86}}{\left(1+\frac{\theta}{2}\right)^3}\cdot\sqrt{\frac{C_f}{2}}\cdot \left(\frac{D}{d}\right) \cdot V_{fB} \cdot V_{fS}\right) \cdot Re_{cr}
\label{eqn_crtitical_Re}
\end{equation}
where $\lambda_B=\frac{\mu_B}{\mu_w}$.

Equation \ref{eqn_crtitical_Re} indicates that smaller $\gamma$ and $\mu_B$ results in smaller $Re_{cr}$, which makes the bitumen liberation happen at lower water velocity \cite{masliyah2011handbook,long2007effect,long2005role}. The change of lump diameter can thus be expressed as
\begin{equation}
\frac{dD}{dt} = -\alpha \left ( MAX\left \{0, \left[ \frac{0.332 \mu_w^2}{\rho_w D^2} (Re_p^{1.5} - Re_{cr}^{1.5}) \right]^\beta\right\}\right)
\label{lump_diameter}
\end{equation}
and the bitumen recovery rate is 
\begin{equation}
\frac{dV_B}{dt} = \frac{\pi}{2}V_{fB}D^2\frac{dD}{dt}
\label{volume_bitumen}
\end{equation} 

$\alpha$ and $\beta$ were determined using CFD simulation results. While the shear stresses may vary across the lump surface, the shear stress at the windward side of the lump is used as the characteristic shear stress since bitumen liberation should be most significant at this spot. 

\subsection{Fitting Parameters in Bitumen Liberation Model}
$\alpha$ and $\beta$ values are found by fitting the model to the bitumen liberation rate in the six CFD simulations. The average bitumen liberation rate, $\frac{{\bar{dV_{LB}}}}{dt}$ is calculated by dividing the total volume of liberated bitumen $V_{LB}$ by the time elapsed. $V_{LB}$ is found by 
\begin{equation}
V_{LB}=V_{B0} \left(1-\frac{V_{vavfB}}{V_{vavfB0}}\right)
\end{equation}
where $V_{B0}$ is the initial volume occupied by bitumen in the lump, $V_{vavfB0}$ is the initial volume average volume fraction of bitumen and, $V_{vavfB}$ is the volume average volume fraction of bitumen at the target time. The simulation results are listed in Table \ref{tab:alpha}

\begin{table}[]

    \centering
    \caption{Conditions and results of six CFD simulations}
    \label{tab:alpha}

\begin{centering}
\begin{tabular}{l l l l l l l}

\hline
Case number & $Re_p$ & FSP & D (cm) &  $Re_{cr}$ \\
\hline
case 1 & 2000 & 36 & 5 & 1565 \\
case 2 & 3000 & 36 & 5 & 1565 \\
case 3 & 4000 & 36 & 5 & 1565 \\
case 4 & 2000 & 16 & 4 & 1252 \\
case 5 & 3000 & 16 & 4 & 1252\\
case 6 & 4000 & 16 & 4 & 1252 \\

\hline

\end{tabular}
\end{centering}
\end{table}

\begin{figure}[htbp]
    \centering
    \includegraphics[width=1\textwidth]{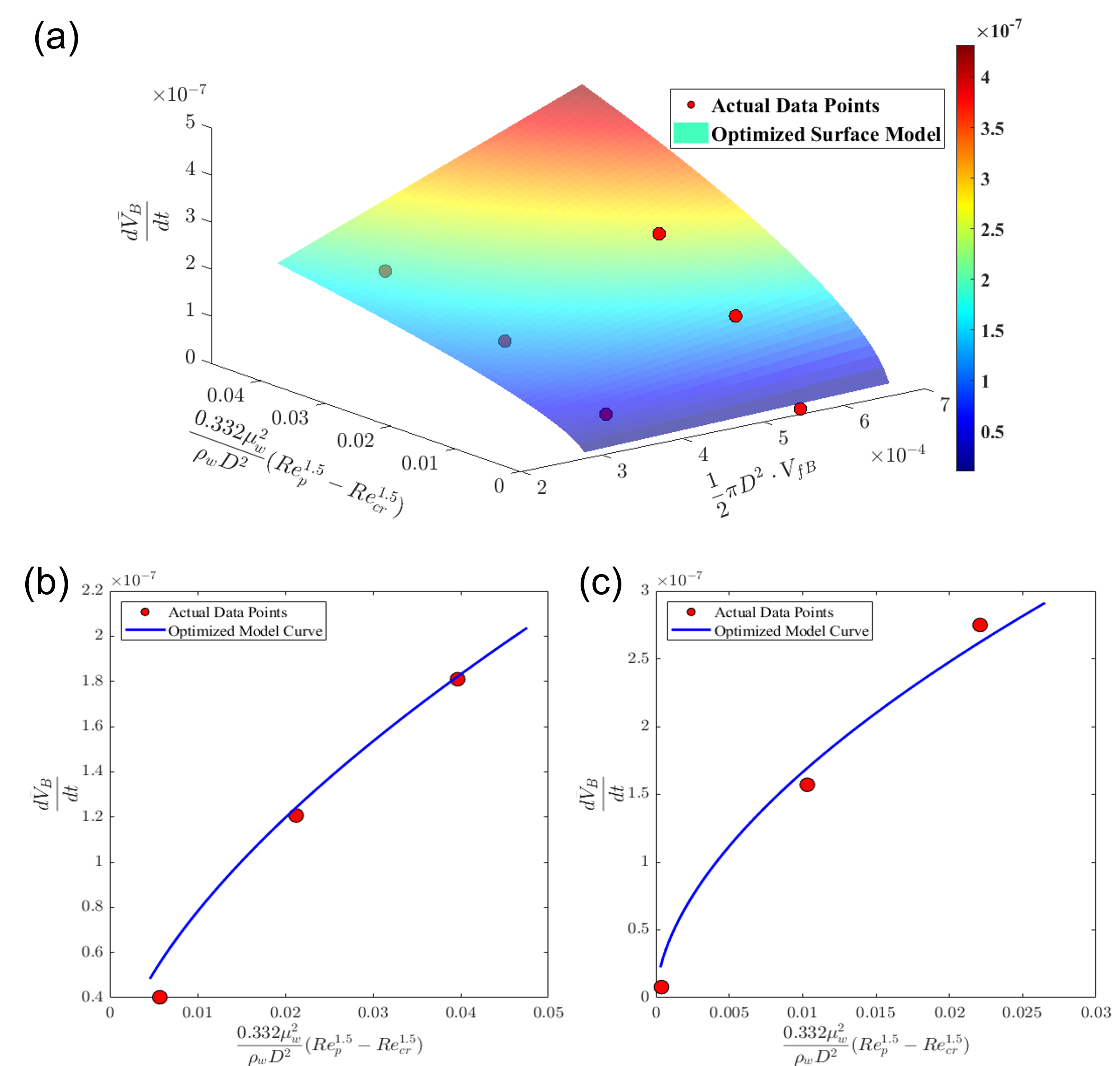}
    \caption{(a) The best fit surface of the model with respect to the six simulation cases (b) The fitted function for 5 cm lump (c) The fitted function for 4 cm lump.}
    \label{fig_curve}
\end{figure}

The best fit parameter values are $\alpha = 5.71 \times 10^{-3}$ and $\beta = 1.40$ with $R^2=0.952$. Figure \ref{fig_curve}(a-c) shows the fitted results. The magnitude of $\beta$ is greater than 1, which implies an amplified effect of the excessive shear stress over critical threshold value. This may be attributed to the formation of a more corrugated surface as the sand particles are removed from the surface of the lump. The rough surface results in complicated flow on the surface, generating micro-eddies that amplify the effect of shear stresses. Moreover, the rough surface may allow water to penetrate into the lump, which can weaken the internal cohesion of the bitumen matrix and accelerate bitumen liberation \cite{pazouki2013experimental,pazouki2019ablation}. 

\subsection{Validation against SRC Data and Comparison with Masliyah's Model}
In the hydrotransport pipelines as the lumps interact with one another and ablate, $v_{rel}$ and $Re_p$ can change. To consider the change of $Re_p$, a transient one-dimensional computational domain was implemented using MATLAB to simulate the SRC experimental setup. The model simulates a total flow time of 3600 seconds. Oil sands lumps with an initial diameter of 2.54 cm and an initial volume fraction of $\alpha_{lump} = 0.4$ were subjected to a continuous water phase velocity of $v_w = 2$ m/s with initial lump velocity of 2 m/s. The lump-lump interactions would alter the slip velocity and $Re_p$, to capture these collisions, the lump-lump interactions are incorporated by estimating a particle mean free path ($\lambda$). The mean free path for the oil sands lumps is based on the number of oil sands lumps in volume ($n$) and lump diameter ($D$) which can be calculated by equation \ref{eqn:lamda_free}.\cite{hsiau1993kinetic,gidaspow1994multiphase,lun1984kinetic}:
\begin{equation}
    \lambda = \frac{1}{\sqrt{2}\pi n D^2}
    \label{eqn:lamda_free}
\end{equation}
The number of oil sands lumps in volume ($n$) is related to the total volume of lump ($V_{lump,t}$) in the domain with given lump volume fraction as \cite{hsiau1993kinetic,gidaspow1994multiphase}.
\begin{equation}
    n = \frac{6V_{lump,t}}{\pi D^3}
    \label{eqn:lamds_relation}
\end{equation}
As the lumps have the displacement of mean free path, a collision occurs which would alter the lump velocity. The collision time can be calculated as \cite{lun1984kinetic,gidaspow1994multiphase}.
\begin{equation}
    t_{collision} = \frac{\lambda}{|\vec{u}_{lump}|}= \frac{D}{6\sqrt{2} |\vec{u}_{lump}| V_{lump,t}}
    \label{eqn:collision_time}
\end{equation}
It is assumed that for each lump-lump collision, the lump velocity is reduced by an coefficient of restitution of 0.9\cite{gidaspow1994multiphase,Ansys}. As a result, 
\begin{equation}
    \vec{u}_{lump}^{n+1}=0.9 \vec{u}_{lump}^{n}
    \label{eqn:velocity_collision}
\end{equation}
where $\vec{u}_{lump}^{n+1}$ is the lump velocity after n+1 collisions and $\vec{u}_{lump}^{n}$ is the lump velocity after n collisions. Since the $\vec{u}_{lump}$ reduces after each collision, $v_{rel}$ increases and, thus, the drag force increases as \cite{Ansys,wilson2006slurry}
\begin{equation}
    \vec{F}_D = K_{w-lump} (\vec{u}_w - \vec{u}_{lump})
    \label{eqn:dragforce}
\end{equation}

\begin{equation}
    K_{w-lump} = \frac{\pi \rho_w C_D |\vec{u}_w - \vec{u}_{lump}|D^2}{8}
    \label{eqn:dragforce_coeff}
\end{equation}

\begin{equation}
    C_D = \begin{cases} 
        \frac{24}{Re_p} \left(1 + 0.15 Re_p^{0.687}\right) & Re_p \le 1000 \\ 
        0.44 & Re_p > 1000 
    \end{cases}
    \label{eqn:sh_Nau}
\end{equation}
where $K_{w-lump}$ is the momentum exchange coefficient and $C_D$ is the Schiller-Naumann drag coefficient \cite{masliyah2011handbook,Ansys,wilson2006slurry}. 

The increased drag force accelerates the lump during the time between two collisions as
\begin{equation}
  \frac{du_{lump}}{dt} =\frac{3 \rho_w C_D |\mathbf{u}_w - \mathbf{u}_{lump}|^2}{4D\rho_{lump}}  - \frac{ \nabla P}{\rho_{lump}}
    \label{eqn:acceleration_lump2}
\end{equation}
where $\nabla P$ is implemented using the measured pressure gradients in SRC pipelines. Since the lump-lump collisions occur at discrete intervals $t_{collision}$, the simulation utilizes a computational time step $\Delta t=0.01t_{collision}$ to resolve the effect of changing lump velocity.

As the accumulated time reaches $t_{collision}$, a collision occur and the momentum loss defined by Equation \ref{eqn:velocity_collision} is applied at this time step. During the time steps between collisions, the lump diameter ($D$) is reduced due to bitumen liberation process, which is modeled by equation \ref{lump_diameter}. The volume of liberated bitumen is calculated using equation \ref{volume_bitumen} and the percentage of bitumen liberated is calculated by 
\begin{equation}
    Bitumen\ Recovery \% = \frac{vavf_{B}}{n_0V_{lump-0}V_{fB0}}
    \label{eqn:BR}
\end{equation}
where $n_0$ is the initial number of oil sands lumps in the domain, $V_{lump-0}$ is initial volume of the lumps, $V_{fB0}$ is the initial volume fraction of bitumen in the oil sands lumps, and $vavf_{B}$ is the volume average volume fraction of bitumen. For each validation case, the $Re_p$ is calculated and the results are listed in Table \ref{tab:ReCr}. The percentage of bitumen liberated results calculated by the MATLAB simulation are validated against the SRC bitumen recovery experimental data.

\begin{table}[]

    \centering
    \caption{Conditions of MATLAB simulations for validation.}
    \label{tab:ReCr}

\begin{centering}
\begin{tabular}{l l l l}

\hline
MATLAB simulation no. & SRC Exp & Critical Re number & -$\nabla P$ (kpa/m)\\
\hline
1  & Exp 1 & 5355 & 1.038 \\
2  & Exp 2 & 6486 & 1.025 \\
3  & Exp 3 & 3821 & 0.904 \\
4  & Exp 4 & 8153 & 1.023 \\

\hline

\end{tabular}
\end{centering}
\end{table}

\begin{figure}[htbp]
    \centering
    \includegraphics[width=1\textwidth]{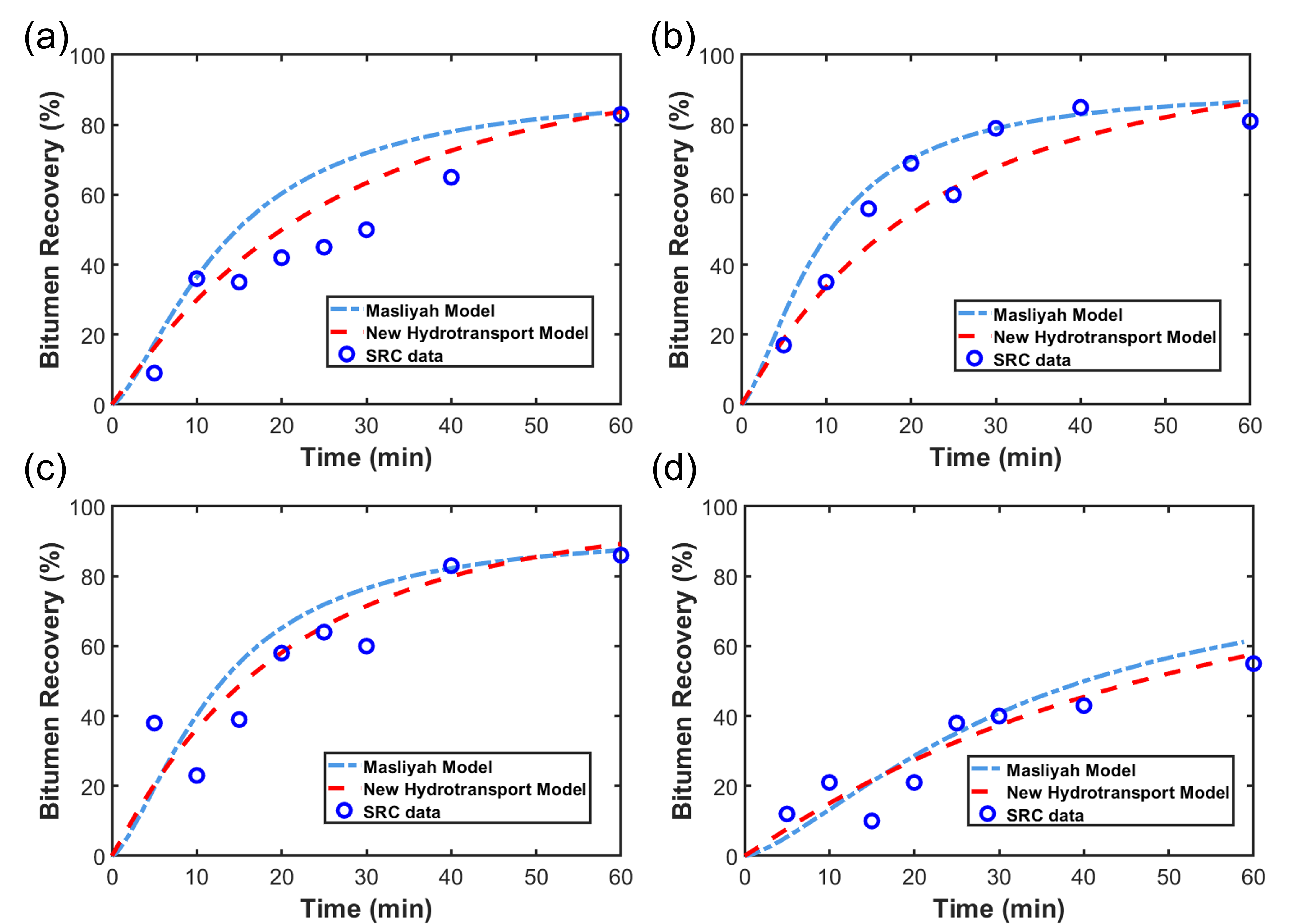}
    \caption{Comparison of proposed shear-driven bitumen ablation model with four experimental cases from SRC and Masliyah's model \cite{masliyah2011handbook,pazouki2019ablation} for (a) Exp 1 (b) Exp 3 (c) Exp 2 and (d) Exp 4.}
    \label{fig_validation}
\end{figure}

\begin{figure}[htbp]
    \centering
    \includegraphics[width=1.1\textwidth]{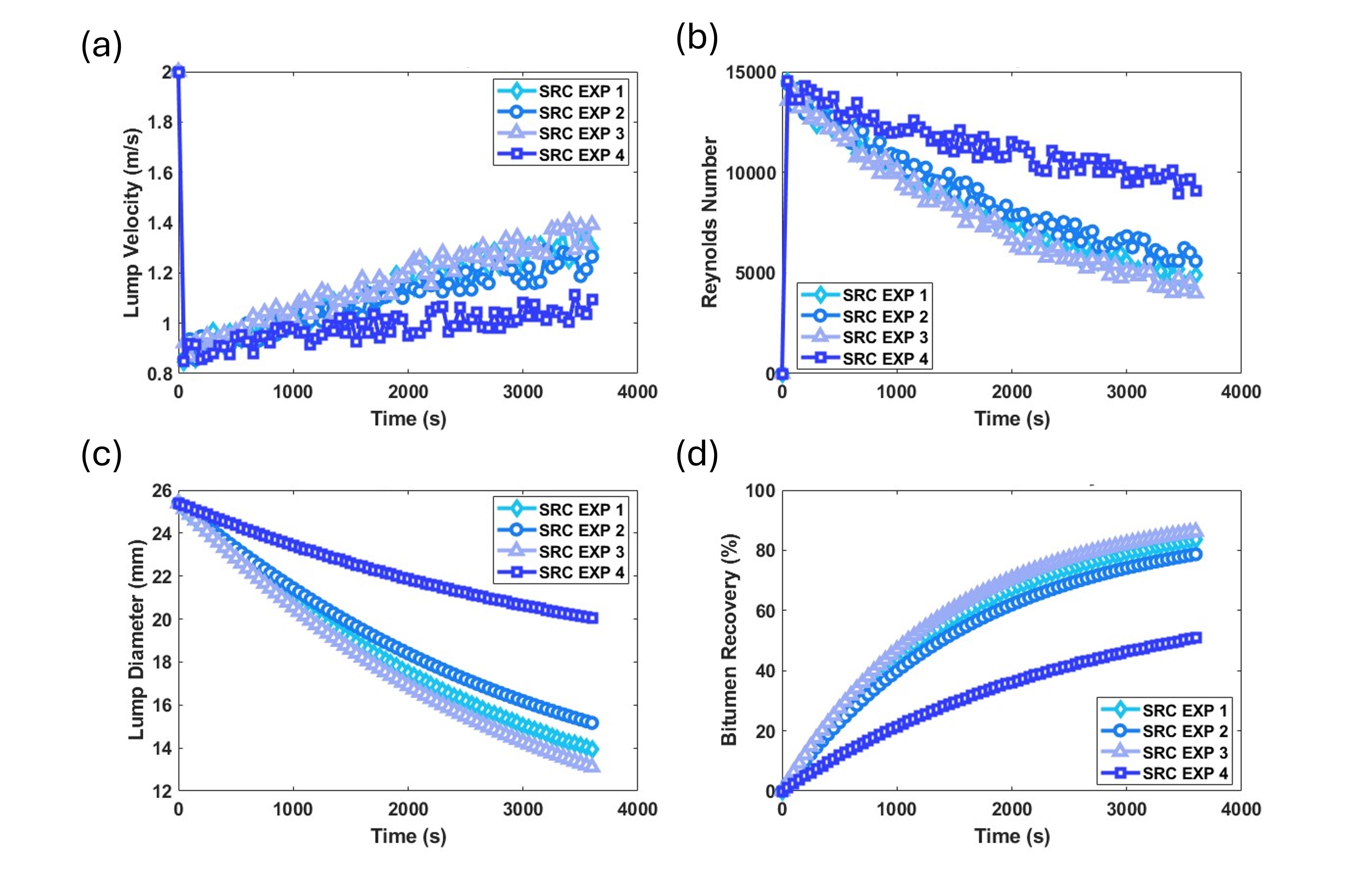}
    \caption{Comparison of MATLAB simulation results (a) $|\vec{u}_{lump}|$, (b) $Re_p$ (c) $D$ and (d) bitumen recovery \% for SRC experimental data sets 1 to 4. }
    \label{fig_validation_2}
\end{figure}

Figure \ref{fig_validation} compares the prediction of the proposed model with experimental results, as well as the prediction by Masliyah model. It shows that the predictions of the proposed model are in good agreement with SRC data with an average error of 8.3\%, which outperforms those of Masliyah model with an average error of 12.1\%. It indicates that integration of drag forces and lump-lump collision mechanics effectively improve the prediction of bitumen liberation rates from oil sands lumps in industrial pipelines. Figure \ref{fig_validation_2} shows the evolution of $|\vec{u}_{lump}|$, $Re_p$, $D$ bitumen liberation \% in the bitumen liberation process. It is shown that due to a conjugate effect of lump collision and increased drag force, $|\vec{u}_{lump}|$ is fluctuating with an overall increasing trend. As a result, $v_{rel}$ decreases over time, which leads to a decreasing $Re_p$ as the lump ablates. Therefore, the rate of decreasing of $D$ is getting slower, which leads to a decreasing rate of growth of bitumen liberation \% at the late stage.

\section{Results and Discussion}
\begin{figure}[htbp]
    \centering
    \includegraphics[width=1\textwidth]{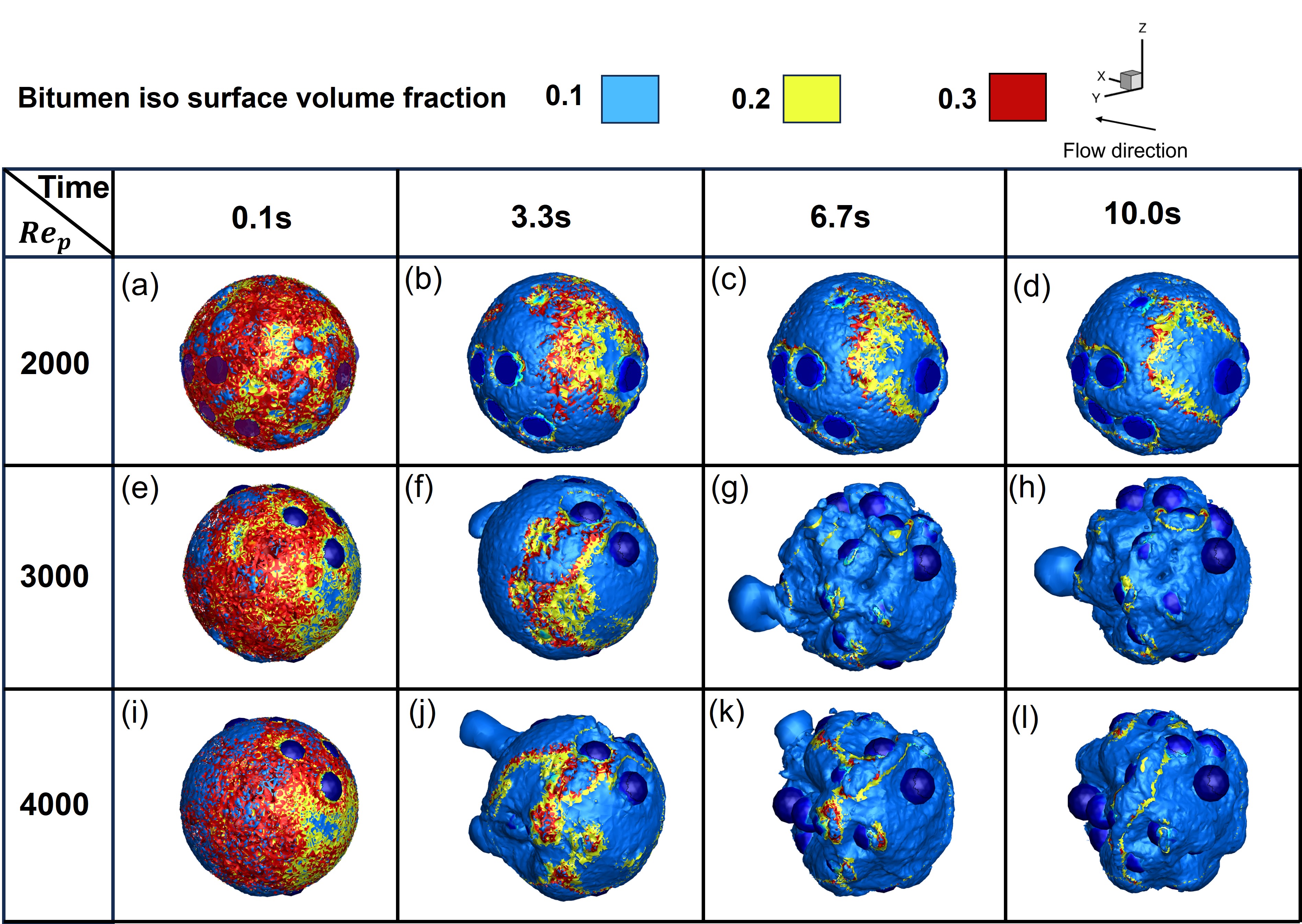}
    \caption{Bitumen volume fraction of 0.3, 0.2 and 0.1 iso surface contours of 5 cm lump for $Re_p$ 2000 at time (a) 0.1s, (b) 3.3s (c) 6.7s (d) 10s, $Re_p$ 3000 at time (e) 0.1s, (f) 3.3s (g) 6.7s (h) 10s, and $Re_p$ 4000 at time (i) 0.1s, (j) 3.3s (k) 6.7s (l) 10s.}
    \label{fig_36p_b_iso}
\end{figure}

\begin{figure}[htbp]
    \centering
    \includegraphics[width=0.8\textwidth]{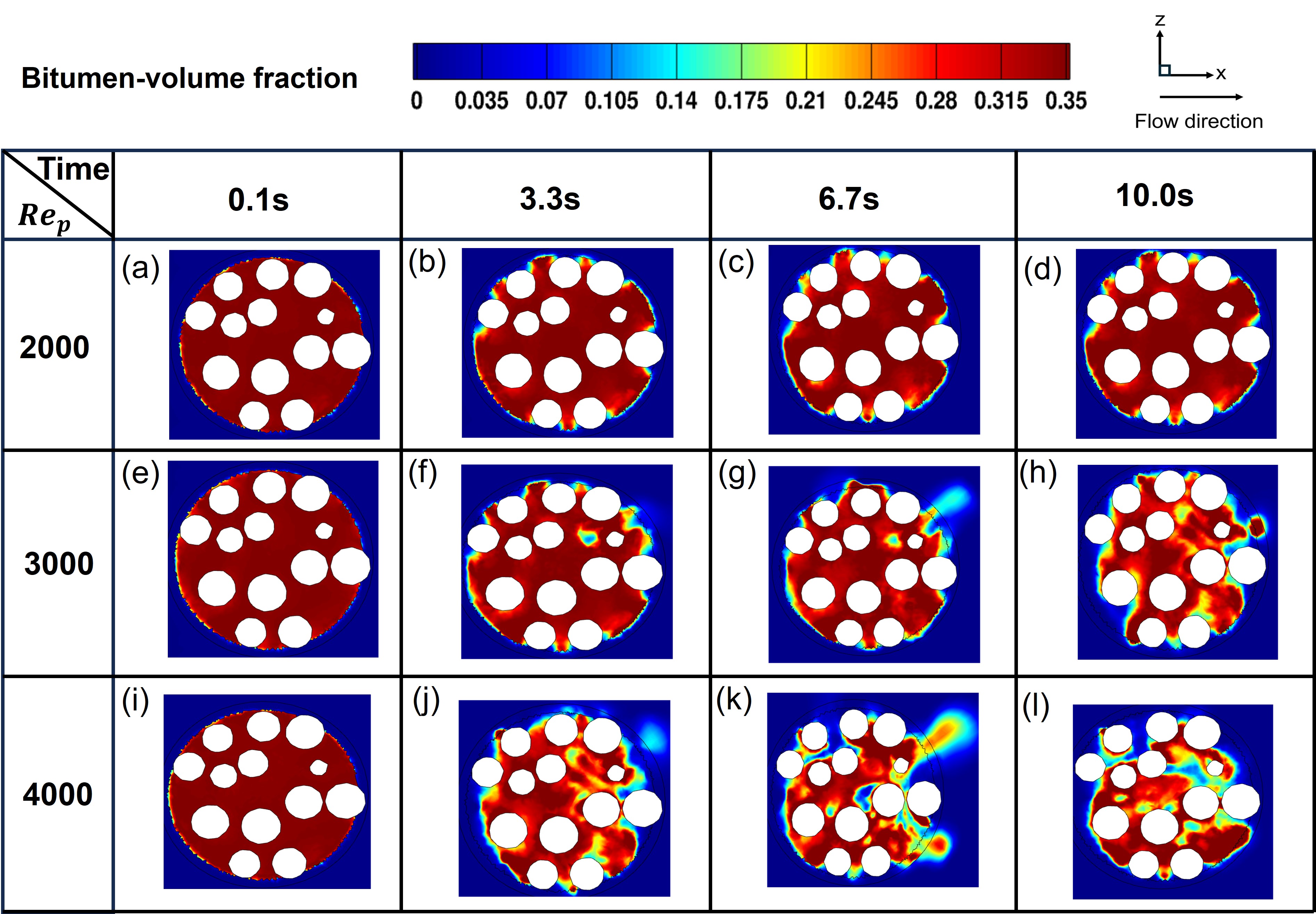}
    \caption{Volume fraction of bitumen at plane $y=0$ of 5 cm lump at t=0.1s for $Re_p$ 2000 at time (a) 0.1s, (b) 3.3s (c) 6.7s (d) 10s, $Re_p$ 3000 at time (e) 0.1s, (f) 3.3s (g) 6.7s (h) 10s, and $Re_p$ 4000 at time (i) 0.1s, (j) 3.3s (k) 6.7s (l) 10s.}
    \label{fig_36p_bs_mid}
\end{figure}

Figure \ref{fig_36p_b_iso} and \ref{fig_36p_bs_mid} show evolution of bitumen volume fraction on the iso surfaces and $y=0$ plane inside the lump for the 5 cm lump at three $Re_p$. The $Re_{cr}$ for the 5 cm lump is 1215. As a result, bitumen liberation happens for all the three cases. As $Re_p$ increase from 2000 to 4000, the bitumen is liberated faster. 

At the initial stage of the bitumen liberation, $t<3.3s$, due to the high viscosity of the bitumen, water is forced to flow aside from the pump rather than penetrating into the lump. As the water flow hits the front of the lump first, the high stagnation pressure flattens the bitumen surface and push the bitumen downstream along the surface of the lump, which results in a ring-shape region with high bitumen volume fraction shown in Figure \ref{fig_36p_b_iso} (b), (f) and (j). The deformation of the bitumen results in the exposure of silica particle matrix inside the mixture, which generates a more corrugated surface to facilitate destabilization of the oil sands lump. In addition, the bitumen inside the lump remain almost intact as shown in Figure \ref{fig_36p_bs_mid} (b), (f) and (j).

\begin{figure}[htbp]
    \centering
    \includegraphics[width=1\textwidth]{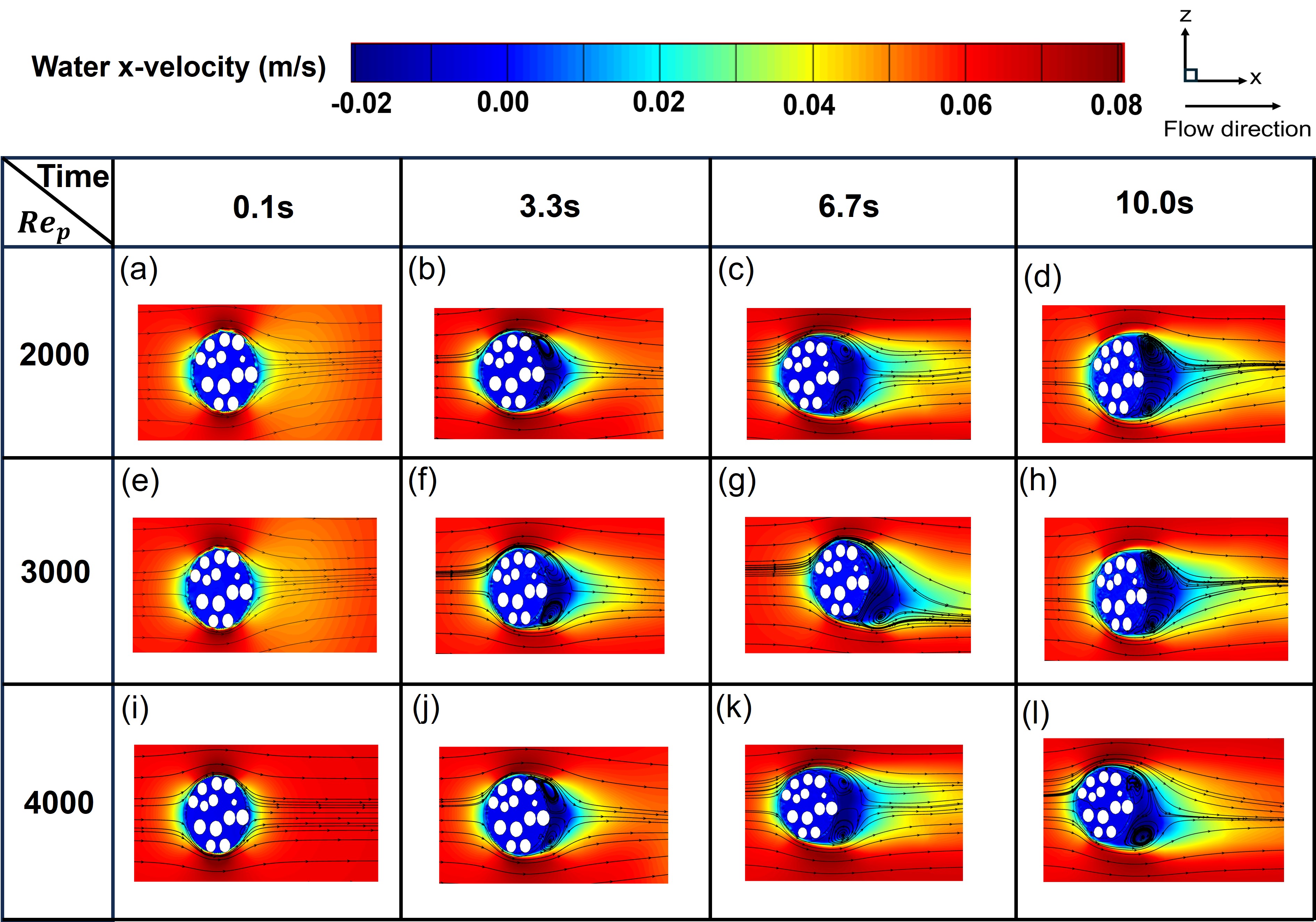}
    \caption{Water x-velocity and streamlines at plane $y=0$ of 5 cm lump for $Re_p$ 2000 at time (a) 0.1s, (b) 3.3s (c) 6.7s (d) 10s, $Re_p$ 3000 at time (e) 0.1s, (f) 3.3s (g) 6.7s (h) 10s, $Re_p$ 4000 at time (i) 0.1s, (j) 3.3s (k) 6.7s (l) 10s.}
    \label{fig_36p_vel}
\end{figure}

As the bitumen liberation proceeds to $t = 6.7 s$ for higher $Re_p$, a small amount of water penetrates into the bitumen due to the reveal of corrugated front face, which is indicated by a reduced volume fraction at the front region of the lump shown in Figure \ref{fig_36p_bs_mid} (c), (g) and (k).  Furthermore, water flow starts to peel off bitumen on the lateral region of the lump where the shear forces is the most significant. Meanwhile, the vortices are formed in the wake region and continuous vortex shedding is observed at the back of the lump as shown in Figure \ref{fig_36p_vel} \cite{caruso2023karman}. These vortices enhances the interaction between water and bitumen, resulting in a faster removal of bitumen at the back of the lump. Such an effect is more significant at higher $Re_p$, where vortices are stronger. It can be visualized by the extended bitumen tail and more exposed FSP at the back of the lump shown in \ref{fig_36p_b_iso} (g) and (k).

At later stage $t = 10 s$, the bitumen removal process slows down significantly. For example, as shown by Figure \ref{fig_36p_b_iso} (l) and \ref{fig_36p_bs_mid}(l), The distribution of bitumen volume fraction in the lump region does not change much from that at $t = 6.7 s$. This may be attributed to the shielding effect of FSP surrounding the remaining bitumen. The fixed FSP results in a narrower pathway filled with highly viscous bitumen, which is hard for water to penetrate. As shown in Figure \ref{fig_36p_vel} (d), (h) and (l), the water velocity inside the lump region, especially between FSP, is much lower compared to that at the surface of the lump. As a result, the shear force exerted by water on the remaining bitumen is weaker and the bitumen liberation process slows down. Such a phenomenon also indicates the increased difficulties to liberate bitumen from the deeper layers of solid particle matrix.

\begin{figure}[htbp]
    \centering
    \includegraphics[width=1\textwidth]{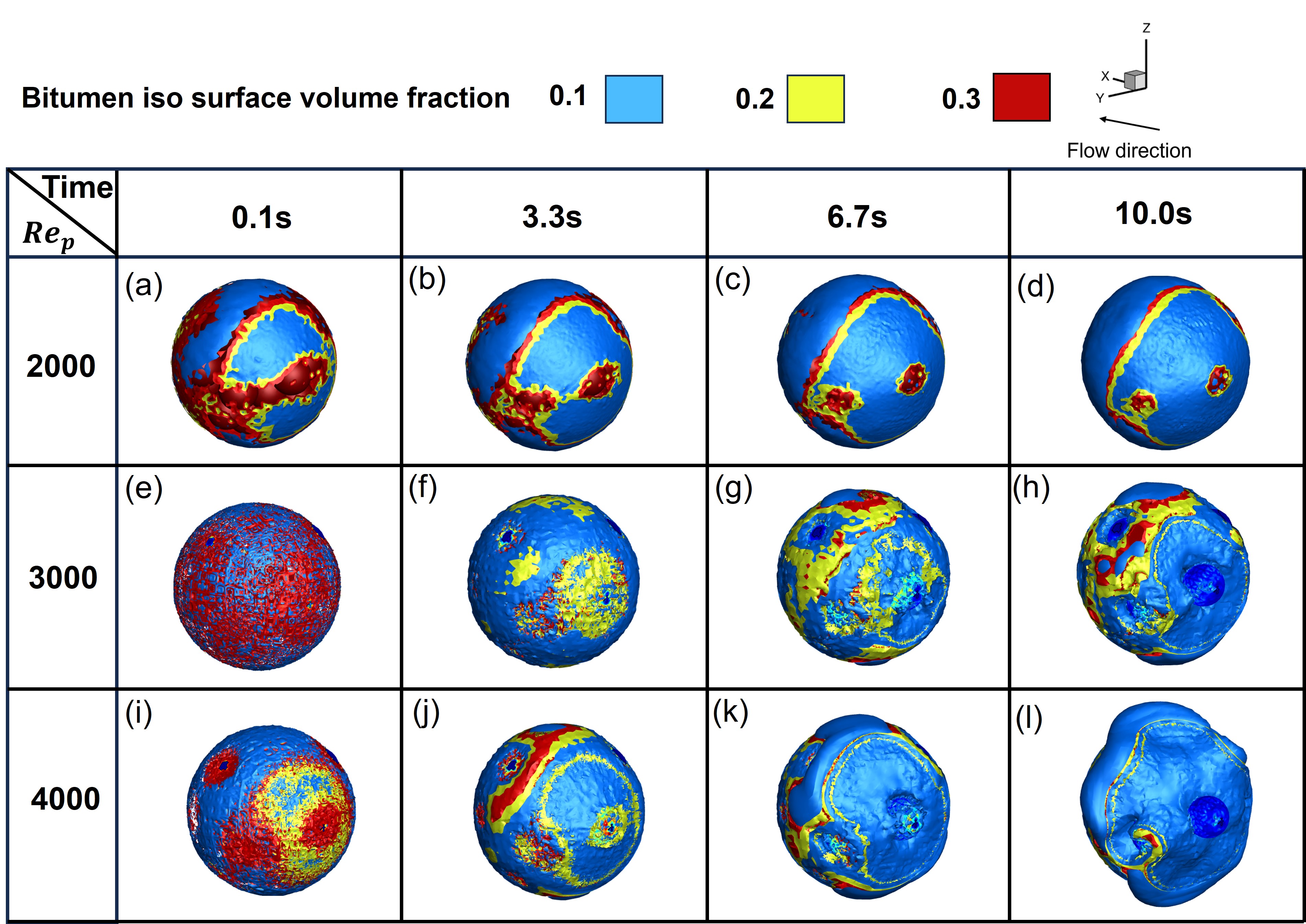}
    \caption{Bitumen volume fraction of 0.3, 0.2 and 0.1 iso surface contours of 4 cm lump for $Re_p$ 2000 at time (a) 0.1s, (b) 3.3s (c) 6.7s (d) 10s, $Re_p$ 3000 at time (e) 0.1s, (f) 3.3s (g) 6.7s (h) 10s, $Re_p$ 4000 at time (i) 0.1s, (j) 3.3s (k) 6.7s (l) 10s.}
    \label{fig_16p_b_iso}
\end{figure}

\begin{figure}[htbp]
    \centering
    \includegraphics[width=0.8\textwidth]{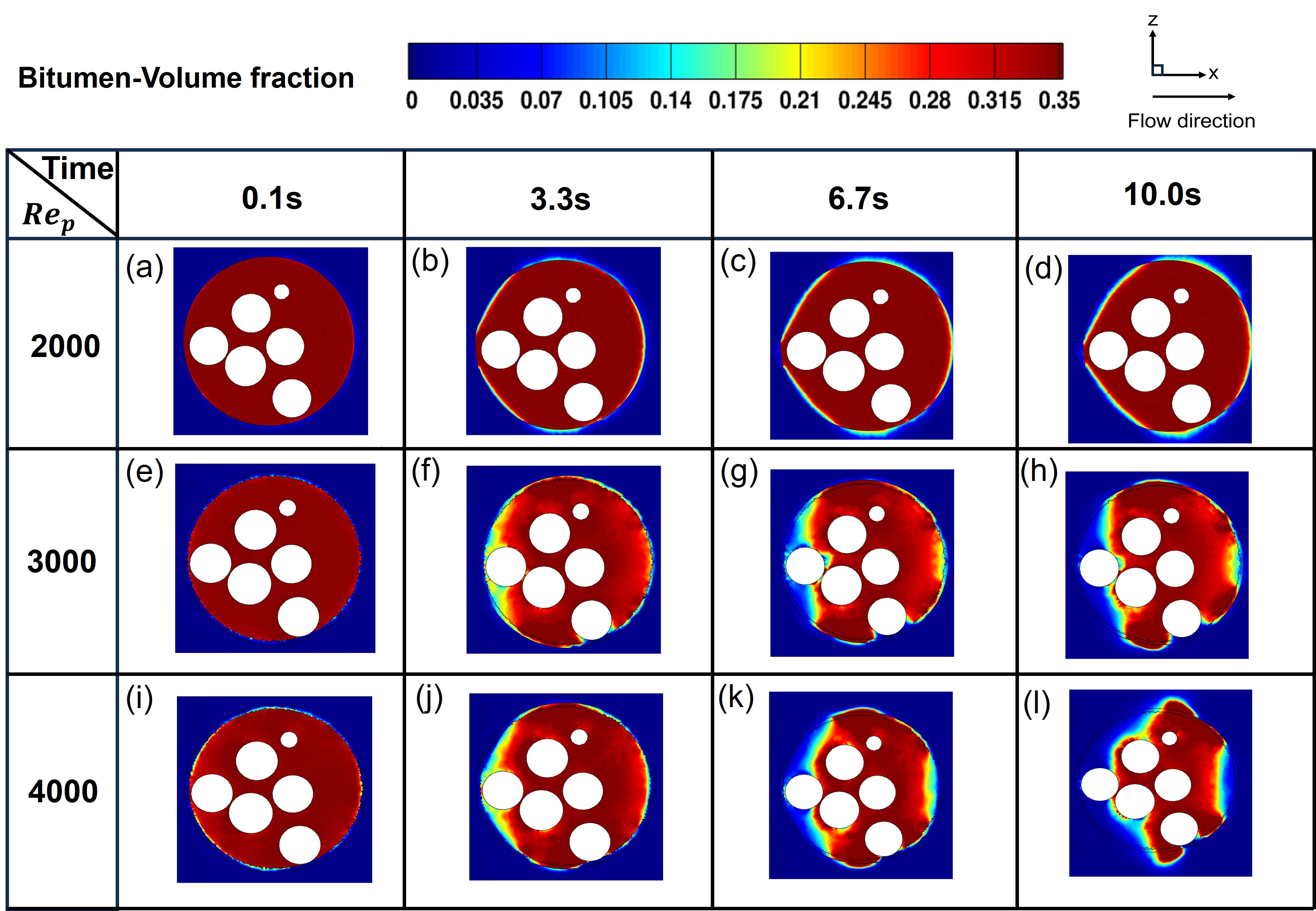}
    \caption{Volume fraction of bitumen at plane $y=0$ of 4 cm lump at t=0.1s for $Re_p$ 2000 at time (a) 0.1s, (b) 3.3s (c) 6.7s (d) 10s, $Re_p$ 3000 at time (e) 0.1s, (f) 3.3s (g) 6.7s (h) 10s, $Re_p$ 4000 at time (i) 0.1s, (j) 3.3s (k) 6.7s (l) 10s.}
    \label{fig_16p_bs_mid}
\end{figure}

Figure \ref{fig_16p_b_iso} and \ref{fig_16p_bs_mid} shows bitumen volume fraction distribution on the iso surface and on $y=0$ plane for the 4 cm lump. The bitumen liberation mechanism and the effect of $Re_p$ on the bitumen liberation process is similar to those for the 5 cm lump, which involves the flatten of the bitumen surface at the front, push back of the bitumen along the surface, and the removal of bitumen due to vortex shedding at the back of the lump. The main difference lies in the number of FSP inside the lump. Due to less FSP in the lump, the bitumen layer is thicker. As a result, the initial removal rate of bitumen is slower, which is indicated by the less deformed bitumen shape shown in Figure \ref{fig_16p_bs_mid} (b), (f), and (j). However, at the later stage, due to the weaker shielding effect of FSP, a larger portion of bitumen is removed simultaneously from the lump, leaving much less bitumen between the FSP as shown in Figure \ref{fig_16p_bs_mid} (g) and (k). This results in an increase in the bitumen liberation rate.

\begin{figure}[htbp]
    \centering
    \includegraphics[width=1\textwidth]{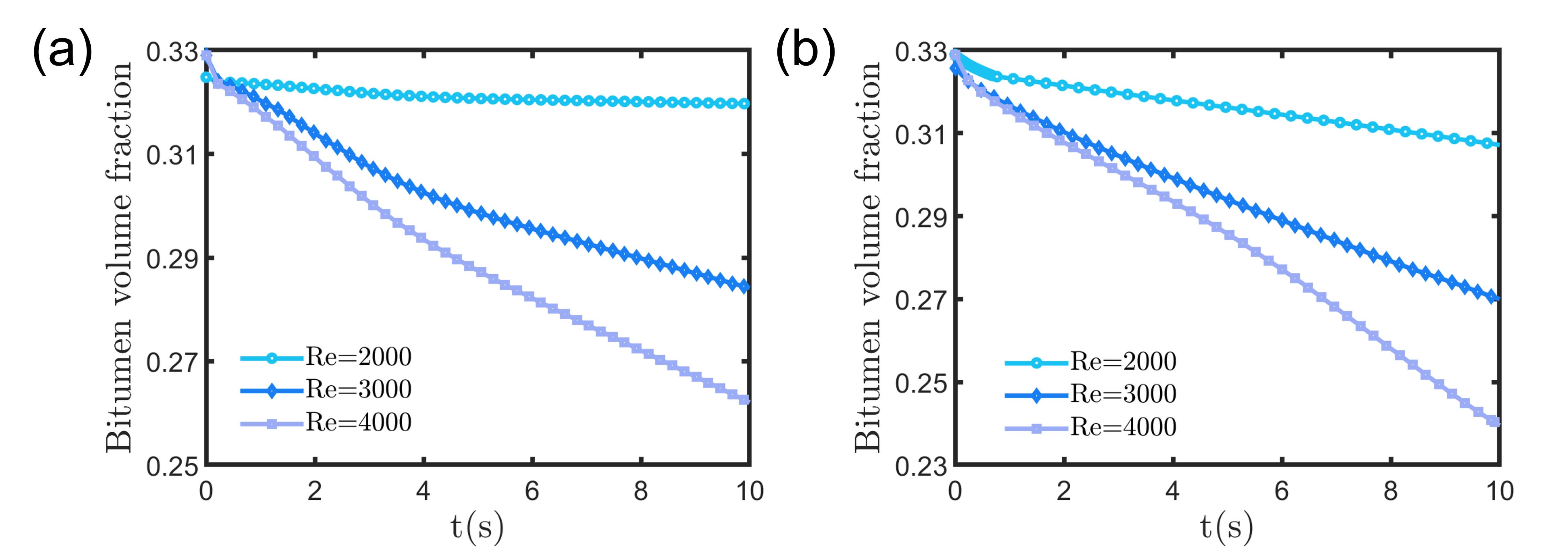}
    \caption{Volume average volume fraction of bitumen within the lump domain for $Re_p$ at 2000, 3000, and 4000 for (a) 5cm lump and (b) 4cm lump.}
    \label{fig_plot}
\end{figure}

Figure \ref{fig_plot} compares the evolution of bitumen volume fraction in the initial lump region for all the six simulated cases. It is shown that the bitumen liberation rate is very slow at $Re_p=2000$. For example, there is only around 1.6\% bitumen removed within 10 s for the 5 cm lump. The bitumen liberation rate increases as $Re_p$ increases, which provides more excessive shear stress over critical value. For example, 13.7\% and 20.4\% bitumen is removed within 10 s from 5 cm lump with $Re_p = 3000$ and $4000$, respectively. Similarly, 6.3\%, 18.2\% and 28.3\% bitumen is removed within 10 s for the 4 cm lump with $Re_p = 2000$, $3000$ and $4000$, respectively. Furthermore, the bitumen liberation rate is faster for the larger lump initially (at around $t<4s$) compared to the smaller lump. This may be attributed to the larger surface area for the larger lump, which makes more bitumen to be removed from the surface of the lump. However, the bitumen liberation rate slows down by 30\% at $t>4s$ due to the shielding effect of larger number of FSP inside the lump. Comparatively, the bitumen liberation rate gets faster for smaller lump due to weaker shielding effect and a consequent removal of larger chunk of bitumen from the lump.

\section{Conclusion}

This study established a 3D Eulerian–Eulerian CFD framework alongside a novel semi-empirical shear-driven model to investigate bitumen liberation dynamics in hydrotransport pipelines. The numerical results demonstrate that the shear forces exerted by the water flow on the lump surface has to be greater than a critical threshold value to overcome the cohesion force such that the bitumen can be detached from the lump. The bitumen removal follows a distinct multi-stage progression. Initially, high shear forces strip the outermost bitumen layer, while leeward vortex shedding drives detachment along the wake of the lump. As liberation proceeds, bitumen trapped deeper within the solid matrix is exposed to water flow and removed at a significantly slower rate, primarily because inner silica particles shield the interface and cause local flow deceleration. Mechanistically, This multi-stage behavior also explains the impact of lump scale: although larger lumps exhibit higher initial liberation rates due to their greater exposed surface area, their dense internal particle packing ultimately restricts fluid penetration, leading to a lower overall extraction efficiency. For example, for $Re_p=4000$, the amount of bitumen liberated within 10 s is reduced by 28\% for a 5 cm lump compared to a 4 cm lump. 

Leveraging these hydrodynamic insights, a semi-empirical model was formulated by balancing fluid drag forces against the interfacial tension and viscous resistance of bitumen bridges between silica sands. The bitumen liberation rate is assumed to be proportional to the excessive shear stress provided by the water flow. By explicitly coupling detachment kinetics to the excess particle Reynolds number beyond the critical threshold, the model offers a physics-based approach to predict extraction behavior, which is validated against SRC experimental data with an error of 8.2\%. Overall, these findings advance the fundamental understanding of fluid-particle interactions in complex slurry and provide a practical, field-validated tool for optimizing hydrotransport pipeline operations and maximizing bitumen recovery.

\section*{Acknowledgment}

This work was supported by Imperial Oil Resources Limited and Alberta Innovates through the Institute for Oil Sands Innovation at the University of Alberta (IOSI) [Project IOSI 2023-09] and Natural Sciences and Engineering Research Council of Canada (NSERC). This research was partially funded by the Canada Research Chairs Program (CRC-2018-00344). We also thank Compute Canada (www.computecanada.ca) for their ongoing support and substantial access to the HPC Cedar and Graham clusters. During the preparation of this work, the authors utilized generative AI tools (Gemini) to assist with language polishing, stylistic refinement, and improving text clarity. After using this tool, the authors reviewed and edited the content as needed and take full responsibility for the content.

\section*{CREDIT AUTHOR STATEMENT}

Mohammadhossein Golchin: Formal analysis (lead); Investigation (lead); Methodology (lead); Software (lead); Validation (lead); Visualization (lead); Writing – review \& editing (equal). Siyu Chen: Formal analysis (equal); Writing – review \& editing (equal). Payman Esmaeili:  Experimental data (lead). Petr Nikrityuk: Methodology (equal); Writing – review \& editing (supporting); Xuehua Zhang: Conceptualization (lead); Funding acquisition (lead); Methodology (equal); Project administration (lead); Resources (lead); Supervision (lead); Writing – review \& editing (equal). 

\section*{Nomenclature}

\begin{longtable}{l p{12cm}}
$A$ &  Surface area (L$^{2}$) \\
$BP$ & Blockage percentage (--) \\
$Ca $ & Capillary number (--) \\
$C_{D} \& C_{d}$ & Drag coefficient (--) \\
$C_{f} $ & Skin friction coefficient (--) \\
$C_{\mathrm{fr}}$ & Friction coefficient between solid phases (--) \\
$D$ & Oil sands lump diameter (L) \\
$d$ & Silica particle diameter (L) \\
$d_p$ & Particle diameter (L) \\
$e$ & Restitution coefficient (--) \\
$f$ & Darcy friction coefficient (--) \\
$F$ & Force per unit volume (M L$^{-2}$ T$^{-2}$) \\
$F_{l,sil}$ & Interaction force between liquid and silica phase (M L$^{-2}$ T$^{-2}$) \\
$g$ & Gravitational acceleration (L T$^{-2}$) \\
$g_0$ & Radial distribution function (--) \\
$h$ & Bitumen bridge lenght (L)\\
$h_{cr}$ & Critical bitumen bridge lenght (L)\\
$L$ & Length (L)\\ 
$l_{d}$ & Distance between two successive layers of oil sands lump (L)\\
$L_{N}$ & Circumference of the silica sands on the oil sands lump surface (L)\\
$L_{NA}$ & Active circumference of the silica sands on the oil sands lump surface (L)\\
$I$ & Unity matrix (--)\\
$I_{2D}$ & Second invariant of the deviatoric stress tensor (T$^{-1}$) \\
$k$ & Turbulent kinetic energy (L$^{2}$ T$^{-2}$) \\
$K_{w_i}$ & Momentum exchange coefficient of phase $w$ and $i$ (M L$^{-3}$ T$^{-1}$) \\
$La$ & Laplace number (--) \\
$m_B \& m_{bit}$ & Bitumen mass (M) \\
$m_{lump}$ & lump mass (M) \\
$m_{sil}$ & Bitumen mass (M) \\
$n$ & Number of particles (--) \\
$N_{As}$ & Number of bitumen bridge active sites (--)\\
$N_{Bs}$ & Number of bitumen bridge sites (--)\\
$N_{ss}$ & Projected number of the silica sands on the oil sands lump surface (--)\\
$p, P$ & Pressure (M L$^{-1}$ T$^{-2}$) \\
$P_{k}$ & Production of turbulent kinetic energy (L$^{2}$ T$^{-3}$) \\
$p_{s_i}$ & Solid pressure (M L$^{-1}$ T$^{-2}$) \\
$p_{\infty}$ & Free stream fluid pressure (M L$^{-1}$ T$^{-2}$) \\
$R$ & Silica particle radius (L) \\
$Re$ & Reynolds number (--) \\
$Re_p$ & Oil sands lump Reynolds number (--) \\
$Re_{cr}$ & Critical Reynolds number (--) \\
$t$ & Time (T) \\
$u, U, v$ & Velocity (L T$^{-1}$) \\
$U_{\infty}$ & Free stream velocity (L T$^{-1}$) \\
$u_{*}$ & Skin velocity (L T$^{-1}$) \\
$u_w$ & Water velocity (L T$^{-1}$) \\
$u_{lump}$ & Oil sands lump velocity (L T$^{-1}$) \\
$v_{bit}$ & Bitumen velocity (L T$^{-1}$) \\
$v_{rel}$ & Lump relative velocity (free stream velocity minus lump velocity) (L T$^{-1}$) \\
$v_{cr}$ & Critical velocity (L T$^{-1}$) \\
$v_{sep}$ & Bridge separation velocity (L T$^{-1}$) \\
$v_{sil}$ & Silica velocity (L T$^{-1}$) \\
$V_{B}$ &  Bitumen volume  ($L^3$) \\
$V_{B0}$ & Initial bitumen volume  ($L^3$) \\
$V_{Bridge}$ &  Bitumen bridge volume  ($L^3$) \\
$V_{fB}$ & Bitumen volume fraction (--) \\
$V_{f-Lump}$ & Oil sands lump volume fraction (--) \\
$V_{fS}$ & Silica volume fraction (--) \\
$V_{LB}$ &  Liberated bitumen volume  ($L^3$) \\
$\Bar{V}_{LB}$ &  Average liberated bitumen volume  ($L^3$) \\
$V_{vavfB}$ & Volume average volume fraction of bitumen (--) \\
$V_{vavfB0}$ & Initial volume average volume fraction of bitumen (--) \\
$x$ & X-axis coordinate (L) \\
$y$ & Y-axis coordinate / Distance to nearest wall (L) \\
$z$ & Z-axis coordinate (L)\\
\\
\textit{Greek symbols} \\
$\alpha$ & Kinetic coefficient of ablation ($L^{2.4}$ $T^{1.8}$ $M^{-1.4}$) \\
$\alpha_{i}$ & Phase $i$ Locally--averaged volume fraction (--) \\
$\beta$ & Exponential Coefficient of ablation (--) \\
$\beta_{ij}$ & Drag coefficient between granular material $i$ and $j$ (--) \\
$\gamma_{\Theta_{s}}$ & Collisional dissipation of energy (M L$^{2}$ T$^{-2}$) \\
$\gamma$ & Interfacial tension coefficient (M T$^{-2}$) \\
$\epsilon$ & Porosity (--) \\
$\Theta$ & Granular temperature (L$^{2}$ T$^{-2}$) \\
$\theta$ & Contact angle (--) \\
$\lambda$ & Mean free path (L) \\
$\lambda_B$ & Bitumen viscosity ratio to water (--) \\
$\lambda_{si}$ & Bulk viscosity (M L$^{-1}$ T$^{-1}$) \\
$\mu$ & Dynamic viscosity (M L$^{-1}$ T$^{-1}$) \\
$\mu_B$ & Bitumen dynamic viscosity (M L$^{-1}$ T$^{-1}$) \\
$\mu_{Si}$ & Solid $i_{th}$ dynamic viscosity (M L$^{-1}$ T$^{-1}$) \\
$\mu_w$ & Water dynamic viscosity (M L$^{-1}$ T$^{-1}$) \\
$\mu_{\mathrm{eff}}$ & Effective viscosity (M L$^{-1}$ T$^{-1}$) \\
$\mu_{t,m}$ & Turbulence viscosity (M L$^{-1}$ T$^{-1}$) \\
$\nu$ & Molecular kinematic viscosity (L$^{2}$ T$^{-1}$) \\
$\nu_{T}$ & Kinematic turbulent viscosity (L$^{2}$ T$^{-1}$) \\
$\rho$ & Density (M L$^{-3}$) \\
$\sigma$ & Interfacial tension coefficient (M T$^{-2}$) \\
$\sigma_k, \sigma_\omega$ & Turbulent Prandtl numbers (--) \\
$\tau$ & Shear stress (M L$^{-1}$ T$^{-2}$) \\
$\phi, \phi_p$ & Volume fraction / Solid packing fraction (--) \\
$\phi_{ls}$ & Energy exchange between fluid and solid phases (M L$^{-1}$ T$^{-3}$) \\
$\Phi$ & Maximum packing fraction (--)\\
$\varphi$ & Bitumen bridge filling angle \& Angle of internal friction (--)\\
$\omega$ & Specific dissipation rate (T$^{-1}$) \\
$\omega_{0}$ & Angular frequency (T$^{-1}$) \\
$\Omega_{k}$ & Angular velocity of pipeline rotation (T$^{-1}$) \\
\\
\textit{Subscripts} \\
$0$ & Initial \\
$bit \& B$ & Bitumen \\
$bridge$ & Bitumen bridge\\
$cr$ & Critical \\
$col$ & Collisional \\
$drag \& d$ & Drag \\
$fr$ & Frictional \\
$i, j, k$ & Phase or directional indices \\
$kin$ & Kinetic \\
$p$ & Particle \\
$rel$ & Relative \\
$\sigma$ & Interfacial-tension \\
$s$ & Solid \\
$si$ & $i^{th}$ solid phase \\
$sil$ & Silica \\
$td$ & Turbulent dispersion \\
$w$ & Water \\
$\infty$ & Free stream \\

\end{longtable}

\newpage

\bibliography{ref.bib}

\end{document}